\documentclass[onecolumn]{aastex631}
\usepackage{amsmath,amstext}
\usepackage[T1]{fontenc}
\usepackage{apjfonts} 
\usepackage[figure,figure*]{hypcap}
\usepackage{CJKutf8}
\usepackage{longtable}
\usepackage{comment}
\usepackage{hyperref}
\usepackage{rotating}
\usepackage{makecell}
\shortauthors{Boley et al.}
\usepackage{xspace}
\usepackage{graphicx, xcolor, url, natbib} % Required for inserting images

\shorttitle{Rings in the Sky}
\shortauthors{Boley, Lawler \& Rein}

\begin{document}

\title{Rings in the Sky: Orbital Data Centres and Potential Impacts to Astronomy and the Sky}

\correspondingauthor{A.~C. Boley}
\email{aaron.boley@ubc.ca}

\
\author[0000-0002-0574-4418]{Aaron C.~Boley}
\affiliation{Department of Physics and Astronomy, University of British Columbia, Vancouver, British Columbia V6T 1Z1, Canada}
\author[0000-0001-5368-386X]{Samantha~M. Lawler}
\affiliation{Campion College and the Department of Physics, University of Regina, Regina, Saskatchewan S4S 0A2, Canada}
\author[0000-0003-1927-731X]{Hanno Rein}
\affiliation{Dept. of Physical and Environmental Sciences, University of Toronto at Scarborough, Toronto, Ontario, M1C 1A4, Canada}
\affiliation{Dept. of Astronomy and Astrophysics, University of Toronto, Toronto, Ontario, M5S 3H4, Canada}

\begin{abstract}

Megaconstellations of orbital data centres (ODCs) have the potential to fundamentally change the sky without a concerted mitigation effort. 
Potential changes include producing large coherent structures that would be visible during twilight and some portions of the night, as well as produce persistent infrared and radio source regions.
We investigate the potential sky impacts of three ODC designs. These designs are based on company filings with regulatory bodies, supplemented with some additional assumptions for brightness modelling and ODC spacings.
While more companies have proposed to launch megaconstellations of ODCs, other potential operators are expected to use the same design principles as presented here.
These systems would produce large ring structures with each ring passing through the sky twice a day.
The largest impacts are expected to occur during winter, where ring structures could be seen sweeping through otherwise dark skies.
The confinement of the rings' orbital nodes  will play a large role in determining whether the rings will be dense arcs in the sky or will produce sky-wide interference.
The rings will persist throughout the day in polar regions, with the potential to interfere with polar science initiatives. 
Such structures will further have societal implications for high-latitude communities.
We use simplified brightness models, which compare well with more detailed calculations, to show that brightness mitigation and/or hard limits on satellite numbers will be essential for preventing a future that has more visible satellites in the sky than visible stars during certain times of night.

\end{abstract}


\section{Introduction}

Initial studies of the impacts of megaconstellations\footnote{
The term ``megaconstellation'' itself refers to an approach to satellite constellation design and operations involving large numbers of satellites and a high cadence of satellite replacement. 
The exact number of satellites is unimportant. 
The term ``mega'', which means ``big'', is used in the same way as it is used in other scientific terms, such as megalodon, megalith, megafauna, and megalomania. 
The use of this term is a potentially important step in minimizing shifting baselines, where a constellation of a few thousand satellites may be presented as ``small'' by some when compared with constellations of many tens of thousands of satellites or more. 
}
on the dark and quiet sky \citep{Hall2021, Walker2021} included exploring the possibility of Earth orbits containing a 100,000 satellites or more \citep{Venkatesan2020}. 
Some of those discussions further warned of a sky with as many visible satellites as there are stars visible to the unaided eye. 
Such concerns were often dismissed as being hyperbolic and unfounded,\footnote{This observation is based on the experiences of the authors.} especially given there is a practice of operators filing for more orbits than what will ultimately be used \citep{Falle2023}.

We are now well into the megaconstellation era. 
Mobilization of the astronomical community, such as through the IAU-CPS, American Astronomical Society's COMPASSE, the European Astronomical Society, as well as support from SKAO, ESO, and the UN Office of Outer Space Affairs, has played a large role in beginning to address the impacts of satellite systems on the sky.
Moreover, these efforts are not inherently new -- the IAU and the international community have been persistent in raising awareness about orbital light and spectrum pollution since the dawn of the space age \citep{ByersBoley2023}, including addressing obtrusive space advertising and other forms of light trespass, such as space mirrors \citep{un_obtrusive_2026}.
The satcon reports \citep{Hall2021}, the Dark and Quiet Sky reports \citep{Walker2021}, the recognition of Dark and Quiet Skies at the United Nations Committee on the Peaceful Uses of Outer Space and the United Nations Environment Programme \citep{un_issue_note}, engagement with national regulators and coordination agreements, and discussions with satellite operators have altogether made notable albeit insufficient progress \citep{un_dqs_2026}. 

As part of this process, the IAU has made several recommendations for limiting the impacts of megaconstellations on the dark and quiet sky \citep{iaucps_position}, for both science and society.
Two of those recommendations, relevant to this work, are that operators should endeavour to operate their satellites (1) at altitudes below 600 km and (2) with a magnitude $V > 7$ for altitudes below 550 km and $V> 7 + 2.5 \log_{10}\left(\rm Altitude/550~km\right)$ above 550 km. 
The first recommendation is a compromise, of sorts. 
Satellites will tend to be brighter at lower altitudes. 
However, a lower altitude also helps to limit the satellite's flight through twilight regions. 
Said differently, it helps to keep the satellites in shadow for longer portions of the night, particularly for latitudes closer to the equator.
With that in mind, the altitude threshold is by no means a panacea. 
As we have detailed in \citet{Lawler2022} and \citet{Boley2022}, persistent night sky impacts can remain at high latitudes during summer months for satellite systems with moderate to high inclinations, where the overall geometry can keep satellites in or near twilight throughout the observer's night (which may in practice be astronomical twilight).

The second, higher-altitude portion of the recommendation is discussed in detail in \cite{SATCON1} and \citet{Boley2025}. 
Briefly, the recommended magnitude limit is chosen to ensure that the satellites are below the threshold for observation by the unaided eye, limiting societal impacts. 
The threshold also is chosen to avoid having non-linear detector responses in large-aperture telescopes.
The inclusion of an altitude dependence is to account for the satellite's sky motion.
A fast-moving, bright satellite can have a smaller detector response than a slow-moving, dim satellite. 
As such, within LEO, the higher the altitude the dimmer the satellite should be to avoid detector non-linearities (e.g., ghost trails). 

These recommendations are being discussed extensively within international fora, and attempts have been made to include these recommendations in licensing rules (e.g., EU Space Act). 
Likewise, some industry actors, including operators such as SpaceX and Amazon LEO, have voluntarily incorporated brightness mitigations into their satellites, with a range of success, which we acknowledge.
Unfortunately, satellites remain well above the IAU brightness recommendation \citep{Mallama2025}.
Moreover, as operators move toward larger LEO satellites, some are not intending to meet the IAU recommendation and instead plan to explore operational mitigations (AST SpaceMobile)\footnote{
Such plans were discussed by AST SpaceMobile at the UN-SKAO December 2025 Dark and Quiet Skies meeting. SpaceMobile is currently deploying its 250 planned satellites, some of which are as bright as the brightest stars \citep{Nandakumar2023}, depending on the details of the reflection.
},
while others have not publicized any mitigation efforts (e.g., Guowang and Qianfan).
Some, who once noted they would strive to meet the magnitude 7 threshold\footnote{https://web.archive.org/web/20210304024442/https://www.spacex.com/updates/starlink-update-04-28-2020/index.html}, have since moved away under the premise that meeting that target, in their view, has now become unrealistic \citep[see, e.g.,][]{Foust2025}. 

With all this in mind, one of the most important steps in protecting dark and quiet skies has so far been avoiding a reality of many tens of thousands of satellites. 

But this is quickly changing as more and more operators launch new megaconstellations.
Most concerning, and the focus of this work, are the potential impacts of proposed orbital data centres. 
An insatiable appetite for promotion of generative AI combined with enthusiasm for satellite megaconstellations has created the conditions for a few companies in a few countries to propose a future that could dramatically reshape the night and radio sky for everyone on the planet.
Such systems would require satellite megaconstellations that are extensive in number, mass, size, and launch rate.
Without careful planning (including exercising restraint), they could become some of the most prominent visible structures in the sky.

This work explores the potential impacts of megaconstellations of ODCs, and is complementary to \cite{marcy2026} and \cite{jangidn_etal}.
We specifically focus on dark sky impacts as seen from the surface of Earth. 
Yet, it must be recognized that there is the potential for severe impacts to space-based astronomy, as well.
In addition, there will be serious implications for the orbital environment and Earth's atmosphere from launches and reentries, to mention just a few among many other concerns.

\section{Methods}

\subsection{Sun-synchronous orbits}

Many of the ODC designs use Sun-synchronous orbits for their satellites.
A satellite is Sun-synchronous if its nodal precession period is equal to Earth's orbital period.
This ensures that the given orbital plane will always have the same orientation relative to the Sun, with an initial polar-terminator orbit remaining that way. 
To good approximation, the necessary inclination for circular orbits is given by \citep[e.g.,][]{vallado2007}
\begin{equation} \label{eq:sunsynch}
    i \approx \arccos\left( -\frac{2\omega_p}{3 J_2 n_s}\left(\frac{a}{R_e}\right)^2\right),
\end{equation} 
where $J_2\approx 1.083\times10^{-3}$ is the zonal harmonic containing most of the information for Earth's oblateness, $n_s$ is the satellite's mean motion, $\omega_p$ is the precession angular rate, $a$ is the satellite's semi-major axis, and $R_e$ is Earth's equatorial radius. 
We use this inclination in constructing the Sun-synchronous orbits in our models instead of the notional inclinations in the operator filings for the given satellites.

Sun-synchronous orbits can be fixed at any particular Earth-Sun-satellite geometry.
However, satellites such as ODCs that have very high power and cooling requirements, the Sun-synchronous orbits of choice are the polar-terminator orbits (i.e., the orbit is highly inclined, and always stays near Earth's terminator), which can potentially remain sunlit all the time, all year long. 
One side of the ODC could further persist in shadow, helping to facilitate cooling, depending on satellite design and operation.
This orbital architecture choice confines the longitude of ascending node to always be near the terminator itself, with the Local Time of Ascending Node (LTAN) set to 6 a.m. (leading Earth's orbit) or 6 p.m. (trailing Earth's orbit), with variations up to approximately 1 hr.
With all this in mind, perpetual illumination is not guaranteed, which as we will show is dependent on the details of the orbital configuration (see section~\ref{sec:shade}). 

\subsection{Satellite distributions}

We model and discuss the on-sky effects of three different ODC megaconstellation designs that have been proposed by corporations in FCC filings in the past few months: an ``X-ring'', an ``X-ring'' plus orbital shell, and a single ring, each described below.  

The satellite spatial distributions for each model are based on operator public filings \citep{fcc_sxodc,fcc_sunrise,fcc_stampede}, as well as supplemental information.  
Such supplemental information is necessary because, due to the large number of satellites, the public filings only contain narratives, general descriptions, and a few representative orbits. 
For this reason, our models are further supplemented by the constellation distributions suggested by \cite{mcdowell_website}, which are themselves informed by additional public sources and private (but not confidential) conversations with operators. 
Although orbit overfiling is a common practice \citep{Falle2023}, the extreme numbers being proposed by multiple operators motivates us to take the number of proposed orbits at face value for any single megaconstellation and calculate the consequences.

\textbf{X-ring:}  The Sunrise constellation by Blue Origin offers an illustrative example of how we construct the spatial distributions for our models.
The submitted narrative states ``This system will consist of up to 51,600 satellites operating in circular, Sun-synchronous orbits from 500-1,800 km in altitude, with inclinations between 97 degrees and 104 degrees, with each orbital plane containing approximately 300-1,000 satellites'' \citep[see `Narrative' in][]{fcc_sunrise}. 
The corresponding Draft Schedule S lists their representative orbits with alternating ascending nodes. 
This, along with the desire to maximize sunlight exposure, we infer the intended design is to use Earth trailing and leading nodes with polar-terminator Sun-synchronous orbits. 
The two nodes lead to an ``X-ring'', with alternating nodes for different altitude rings (which would prevent ring intersections). 
The inclination for any given altitude can be further determined by ensuring the precession rate is Sun-synchronous, which is consistent with the representative filings to within a degree or so. 
For the altitude intervals, we use the spacings suggested by McDowell, which themselves are based on the filings when given.

\textbf{X-ring plus orbital shell:} The SpaceX ODC constellation (SXODC) is constructed in a similar manner.  
However, the system has a pseudo-Sun-synchronous design, where a Sun-synchronous X-ring is combined with a dense band of satellites of inclinations approximately $30^\circ$ \citep{fcc_sxodc}.
The recent SpaceX update to their design helped to clarify the intended configuration \citep{spacex_29may2026}, although it also creates some confusion. 
For example, the document's Table 1 lists the total number of satellites as 1 million, but the actual number listed for each orbital configuration sums to 1198120.  

\textbf{Single ring:} As a final constellation design, we include Stampede by Cowboy Space, which is for 20,000 ODCs.
Unlike Sunrise and SpaceX, filings \citep{fcc_stampede} suggest that there is only one node explored, creating a ``single'' Sun-synchronous ring instead of the X-ring configuration. 
Even if other designs are eventually pursued for Stampede, it remains constructive to consider the implications of a single ring system. 

For any of the above constellation architectures, it is unclear how much variation there will be in ring nodal alignments.
Variation in nodes could be pursued for several reasons, including space traffic management. 
We thus present two possibilities for each ODC megaconstellation: (1) a ``tight'' configuration, where we set the nodes to an LTAN of 6 p.m. or 6 a.m. for the X-rings. 
Stampede uses a slightly offset design, with LTAN set to 6:48 a.m. (for reasons we can only speculate).
We also (2) present a ``relaxed'' variation of the nodal configuration, where the nodes of the subrings are spread out by a uniform random distribution between $\pm10^\circ$ of the ``tight'' node configuration.
This choice is informed by the filed tolerances of nodal variations. 
Although larger deviations are included in some filings, pursuing those could also interfere with maximizing solar illumination durations. 

Within any given ring, for either the tight or relaxed configurations, satellites are populated using a random uniform distribution for the mean anomaly. 
In practice, operators will need to exercise station-keeping to maintain safe distances between satellites, although that is an unnecessary complication here for exploring possible sky impacts.
The full constellation architectures used in our models are publicly available.\footnote{For example, see \url{https://github.com/norabolig/odc_sky_impacts}.} A summary is given in Table \ref{tbl:configurations}.

\begin{table}
\begin{center}
\caption{ODC Megaconstellation Design Summary}\label{tbl:configurations}
\begin{tabular}{l c c c c}\hline
ODCs & Configuration & Number & Inclination & Altitude Range (km)\\
Sunrise & X-ring & 51,600 & Sun Sync & 500 to 1800 \\
SXODC & X-ring & 598,720 & Sun Sync & 565 to 1002 \\
 - & Shell & 599,400 & $26^\circ$ to $32^\circ$ & 550 to 978\\
Stampede & Ring & 20,000 & Sun Sync & 700 to 1000\\
\end{tabular}
\end{center}

\end{table}

\subsection{Satellite area}

Operators have provided incomplete information about potential sizes, masses, and power needs of their spacecraft in the initial filings. 
Indeed, SpaceX recently filed an update at the FCC's request with additional information \citep{spacex_29may2026}, including the expected area of their satellites.
Although many questions remain, the expected cross sectional area of the SXODC satellites is listed as being about $800~\rm m^2$, which we use going forward. 
This area includes the satellite, radiators, and solar panels \citep{spacex_29may2026}.\footnote{
This information is evolving quickly, and while  the cross sectional area used above was corroborated earlier in 2026 through statements by SpaceX \citep{LukeJames2026}, SpaceX also recently filed additional information where they note the possibility of using {\it even larger areas on orbit} \citep[1500 m$^2$;][]{spacex_16sep2026}, as did Starcloud \citep[1320 m$^2$;][]{starcloud_17sep2026} and Cowboy \citep[3870 m$^2$;][]{cowboy_18sep2026}.
}
We do not have detailed information for Sunrise, so with an absence of this information, we assume they are essentially the same. 
With that in mind, Cowboy Space provided a sketch of their satellite design in an ex parte filing to the FCC \citep{cowboy_exparte} that implies a solar panel area of approximately 4800 m$^2$.
No size measurements are given for the radiators, but they appear to be similarly sized to the solar panels.  
For reference, the solar panel area on the International Space Station is 2250 m$^2$ \cite{iss_solarpanels}.
Thus, the proposed Cowboy area is so large, that we take the SpaceX size of $800~\rm m^2$ as a relatively conservative size.
Should such sizes on orbit be pursued, then the sky impacts could be much worse than explored here, depending on the detailed satellite design and operations \citep[as an example, see][]{marcy2026}.

Without transparent on-sky brightness modelling published by the operators, it is necessary to make assumptions about satellite properties to estimate the impacts will have on the sky and astronomy. 
It may be easy for some to say that some detail in the modelling process is wrong and therefore the estimates regarding the impacts on astronomy are also entirely wrong, dismissing the predictions. 
Such potential criticism emphasizes (1) the need for transparency. 
It would also (2) ignore many aspects about the model that are correct, with some details being altered but the big picture remaining the same. 
This takes us to how we model the brightness of the satellites.

\subsection{All models are wrong, some are useful: on-sky brightness} \label{sec:brightness_models}

The actual brightness of each satellite will depend on the exact satellite design, including materials used, and the actual operational practice, such as satellite attitudes and deviations. 
Strategies for brightness mitigation will also be important, and could be different between operators. 
Moreover, to our knowledge, the detailed information required to make very high fidelity brightness calculations for each satellite design has not been made public by any megaconstellation operator. 
It also frequently changes.

There are several ways forward. 
One is to focus on a single operator using as much information about the satellite design as practicable, and pair that with bidirectional reflectance distribution functions (BRDF). 
This is the strategy taken by \cite{jangidn_etal}.
Another approach, which we adopt, is to avoid picking a specific satellite model and instead use a simplified approach that can serve as a reference value. 
Here, we use the Lambertian sphere model (LSM), which we previously found was, on average, able to reproduce observed on-sky Starlink brightnesses in 2020-2021 \citep{Boley2022, Lawler2022}.

Both approaches have advantages and limitations. 
Using a specific satellite model and a corresponding BRDF will give the highest fidelity results -- for that specific model and BRDF. 
Changes to either the satellite structural design or the brightness mitigation strategy, both of which  has changed frequently for even an ``established'' megaconstellations like Starlink, could result in large changes to the on-orbit brightness \citep{mallama2026}. 

The LSM, in contrast, does not represent a specific satellite and is highly idealized. 
Yet, it provides a simple and understandable reference, and in some ways, behaves as a type of average. 
Any light pollution/reflection mitigations taken by operators should see the on-sky brightness become much dimmer than that expected for a Lambertian sphere. 
In this way, the LSM brightness is also a useful reference. 

Despite its simplicity, the range in brightness we see using the LSM corresponds, roughly to the range of brightness seen by \cite{jangidn_etal}, who again use a detailed satellite model and BRDF. 
Thus, for understanding possible night sky impacts, the LSM indeed retains substantial value as an overall indicator of possible on-sky light pollution.

The model calculates approximate V band observations.
Thus, the LSM used in these calculations is
\begin{equation}
V = -26.77 -2.5\log_{10} \left(\frac{2\zeta}{3\pi^2} \left( \left(\pi - \phi\right) \cos(\phi) + \sin(\phi) \right) \right) + 5 \log_{10} R + k \chi(Z).\label{eqn:mags}
\end{equation}
Here, $\zeta$ is the combined cross-sectional area and albedo, which we take to be $\zeta\approx 0.2 \times 800~\rm m^2 = 160~\rm m^2$.
This albedo value is based on the results of \cite{Lawler2022} and \cite{Boley2022}, applied to early Starlink models using the LSM.
Should either the albedo or the cross-sectional area change (or both), then the resulting brightness will be affected accordingly.
The phase angle is given by $\phi$, and $k$ is airmass factor for airmass $\chi$ evaluated at zenith angle $Z$. 
We use the \cite{Kasten1989} airmass function with $k=0.15$.
Finally, the satellite range $R$ is included, with units consistent with those used for $\zeta$.

\subsection{Independent Models: Consistent results} 

We use two simulation codes developed by the co-authors independently, reaching consistent results. 
One model uses the same code as that developed for \cite{Lawler2022}, but extended to include the ODC megaconstellation configurations.\footnote{
The code is available here: \url{https://github.com/hannorein/megaconstellations/}.
}
This code incorporates Rebound \citep{Rein2012} to integrate satellite positions along their orbits, giving an animated view of any given configuration.
The other code, while initially built in parallel with the work of \cite{Lawler2022}, was kept separate for independently checking results. 
That code was further extended to include ODC megaconstellations, again independently, and relies on python libraries only.\footnote{
The code is available here: \url{https://github.com/norabolig/odc_sky_impacts}.
}
In both cases, satellites are distributed about Earth according to the given orbital configuration. 

In the python code, the noon-midnight directions are picked in advance to establish the initial orientation. 
The observer is then placed at the desired latitude, and rotated to a specific local solar time. 
Then, the observer and satellites are rotated together, such as to place the Sun on, say, the $-x$ axis in a way that is consistent with the solar declination representing the chosen time of year. 

With the overall orientation now established, satellites are calculated to be sunlit or not for further evaluation.
In the case of the Sun being on the $-x$ axis, satellites with $x\le0$ are sunlit, as are satellites with $x>0$ and $(y^2+z^2)^{1/2} > R_\oplus$.
Here, we take $R_\oplus\approx6378$ km, which is Earth's equatorial radius. 
For these calculations, we assume a spherical Earth, and satellite altitudes are set relative to the given $R_\oplus$.
We discuss some implications of this simplification further in Section~\ref{sec:shade}, but it is not seen as a limitation of this study. 
Moreover, while a spherical Earth approximation causes a shift in some of the satellite positions relative to what an observer would see if placed on Earth's actual geoid, the shifts are small for the purposes here and do not change the overall results.

Using the distribution of sunlit satellites, the vectors relative to the observer are determined, along with solar phase angles and ranges.
Positions are projected onto the local sky by determining the altitude and azimuth of the satellite relative to the local observer.
With this information, the satellite's airmass can be calculated, along with its altitude, according to equation (\ref{eqn:mags}).
For specifics regarding the projections, we refer the reader to the two simulation codes.

\subsection{Throwing Shade} \label{sec:shade}

One immediate issue that becomes apparent when constructing the proposed ODC megaconstellation models is that no single proposed configuration achieves perpetual sunlight for all Sun-synchronous satellites -- a major purported motivating reason for placing ODCs into space.
While this may seem obvious for the SXODC $30^\circ$ inclination band, it is true for all or some of the satellites in the Sun-synchronous rings in each model we explore. 
The reason for this is that each ring's Sun-syncronous precession is about Earth's pole, not the ecliptic plane. 
This means that while a ring may be fully illuminated during one solstice (say June), one polar region of the ring could be in shadow during the other solstice (December), depending on the altitude.\footnote{
See also the discussion by H.~Lewis at \url{https://www.linkedin.com/pulse/sun-always-shines-tv-hugh-lewis-tlkbe/}
} 
Thus each ring will have an eclipse season, with the eclipse period set by whether the node is close to a LTAN of 6 a.m.~or 6 p.m. 
The LTAN will further determine whether the eclipse is the northern or southern sections of a given ring. 

As we will see, the Stampede ring and SXODC X-rings have sections that are fully in shadow because they use too low of altitudes to achieve perpetual illumination. 
The X-ring configuration further ensures that one of the rings always has a section that is in shadow during a solstice. 
The Sunrise megaconstellation uses a wider range of altitudes, so only its lower orbits of its X-rings become eclipsed during solstices. 

The exact altitude for which a polar-terminator Sun-synchronous orbit will have an eclipse during some part of the year is straightforward to calculate for the spherical Earth approximation, where to ensure perpetual illumination the minimum altitude
\begin{equation}
    A_{m} = R_\oplus\left(\sec(\phi_\oplus+i_m(A_m)-90^\circ)-1\right)\rm ,
\end{equation}
with $i_m$ being the corresponding Sun-syncronous inclination and $\phi_\oplus$ Earth's obliquity to the ecliptic.
In practice, Earth is not a perfect sphere and the atmosphere can have nontrivial effects. 
However, the altitude at which we can expect significant scattering of sunlight is similar to the difference in equatorial and polar radii of Earth, making the spherical Earth approximation reasonable.
Here, the minimum altitude $A_m\approx 1400$ km.

An implication of this is that none of these proposed ODCs can be within the coveted region of extreme low-latency altitudes and also receive perpetual illumination for the entire year -- only altitudes above 1400~km receive light for 24 hours a day all year long.
This further means that most of the proposed ODCs will go in and out of shadow at some point of the year every $\sim$90 minutes, during which time they will have to rely on batteries and/or will need to pause major calculations, all while managing large temperature swings. 
While satellites routinely operate in and out of Earth's shadow, the demands set forth by ODCs make eclipses in this context much more challenging. 
Again, ODCs are being sold in part on the idea of perpetual illumination, something that is not achieved by the proposed, public designs.

\section{Results}

\begin{figure}\label{fig:sunrise_Earth_map}
\centering
\hspace*{-0.0cm}\includegraphics[width=6cm]{ 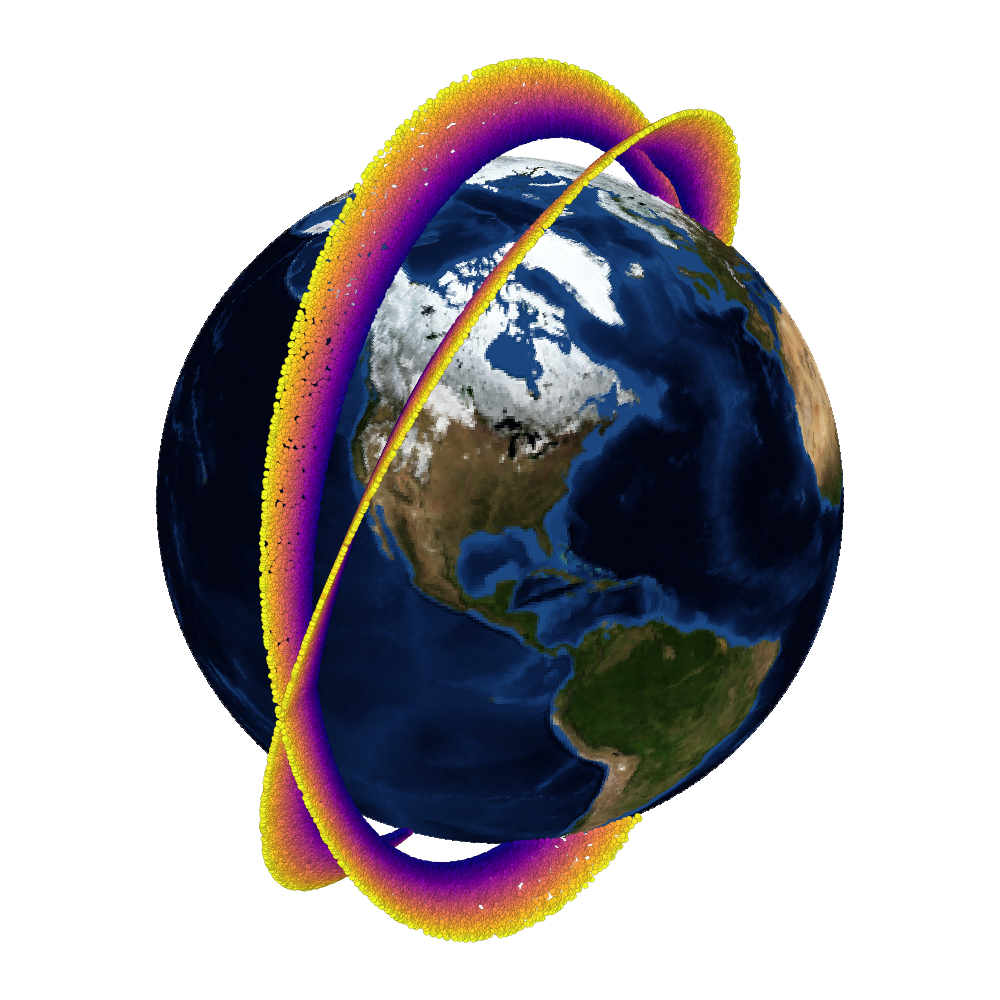}\hspace*{-0.0cm}\includegraphics[width=6cm]{ 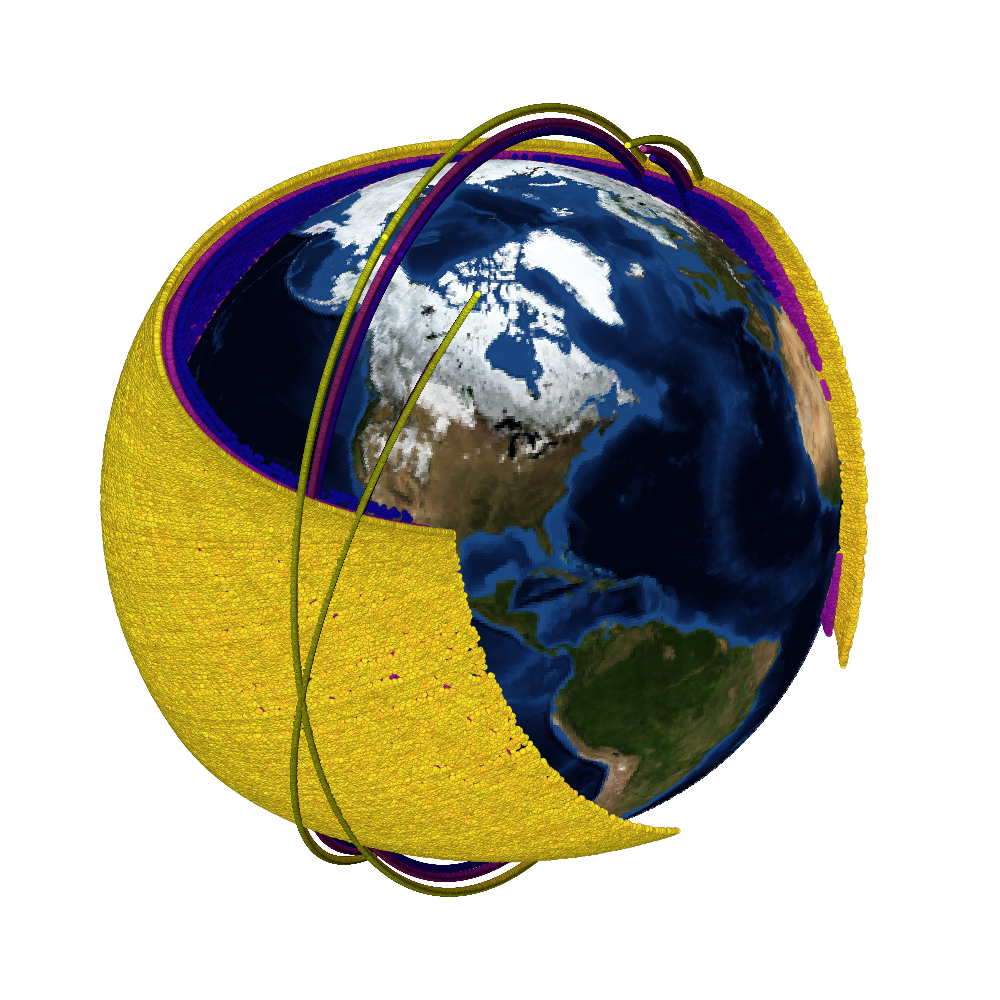}\hspace*{-0.0cm}\includegraphics[width=6cm]{ 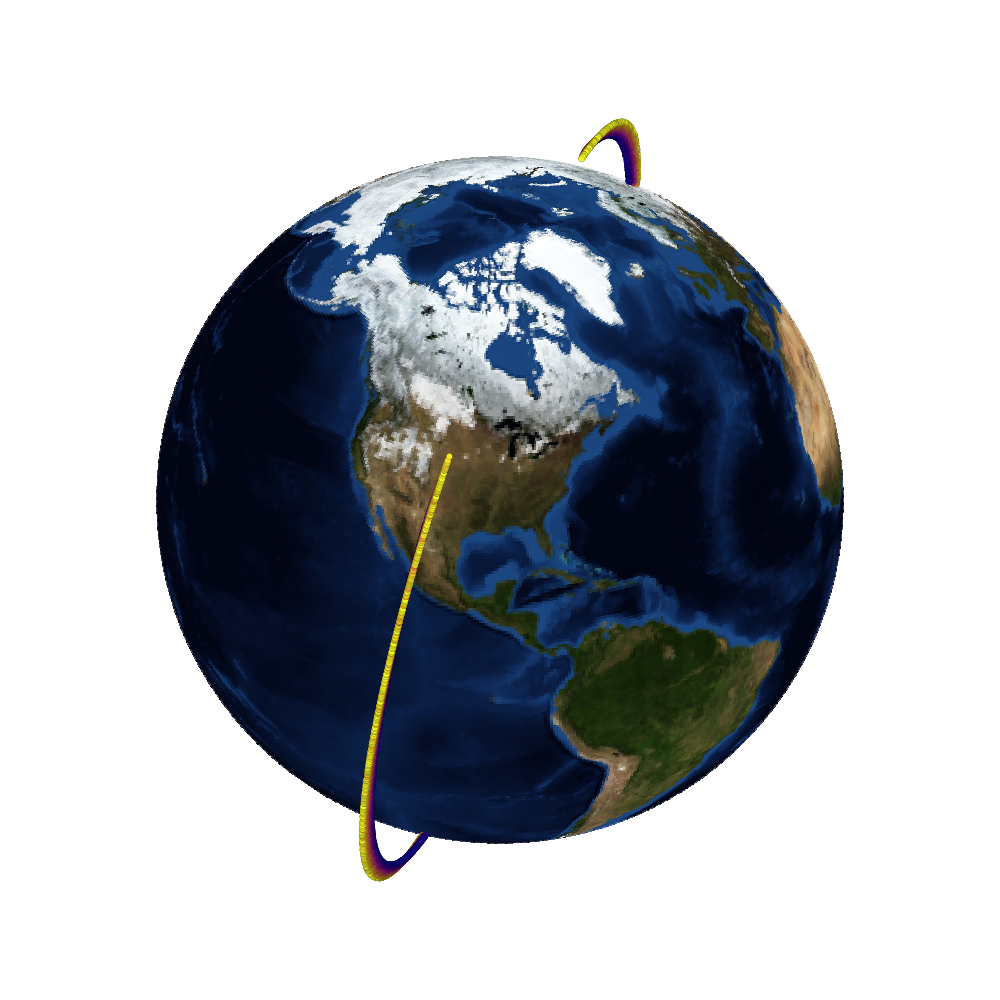}\vspace*{-0cm}
\hspace*{-0.0cm}\includegraphics[width=6cm]{ 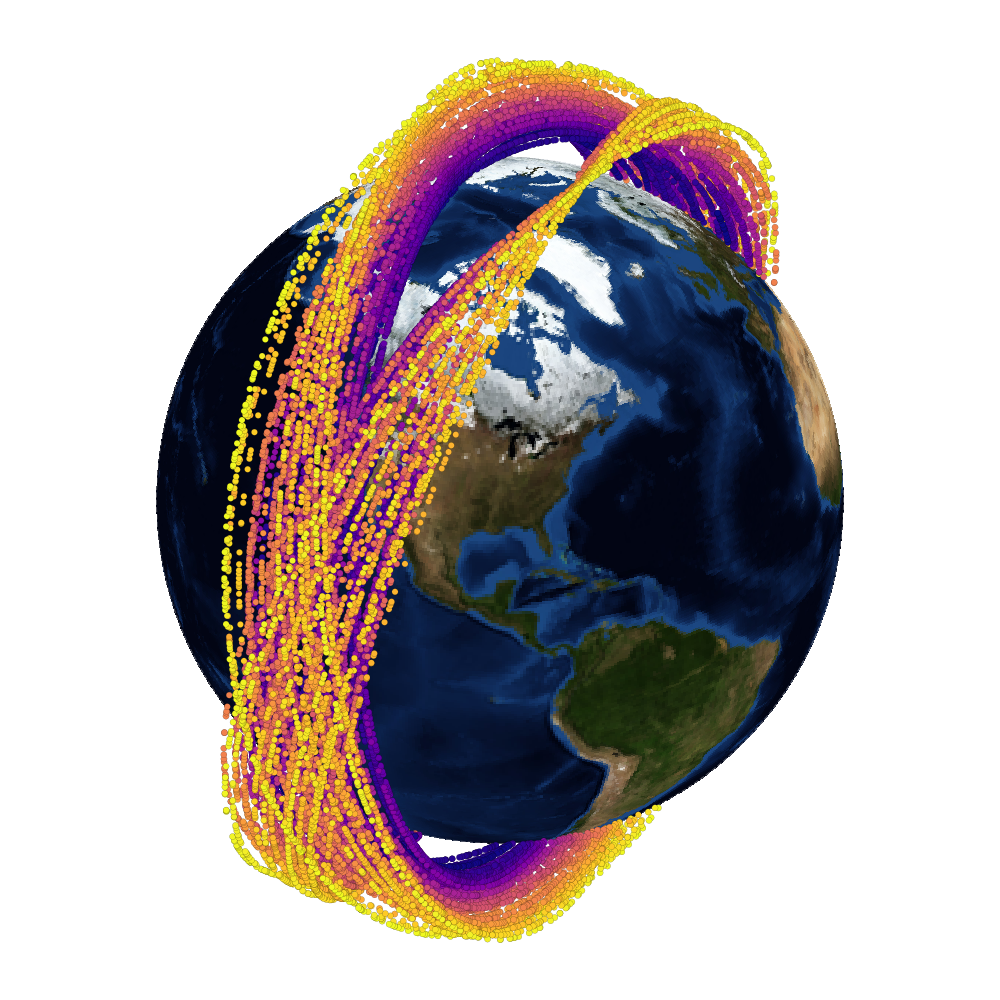}\hspace*{-0.0cm}\includegraphics[width=6cm]{ 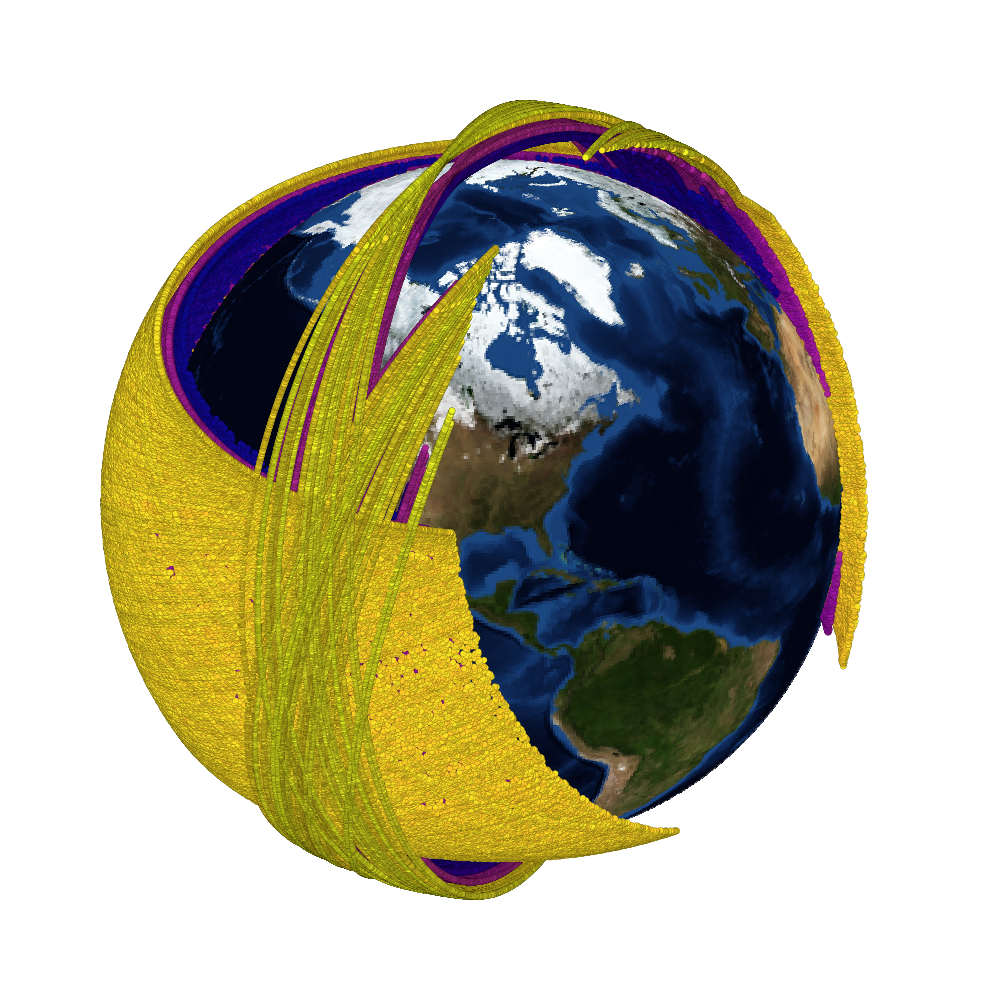}\hspace*{-0.0cm}\includegraphics[width=6cm]{ 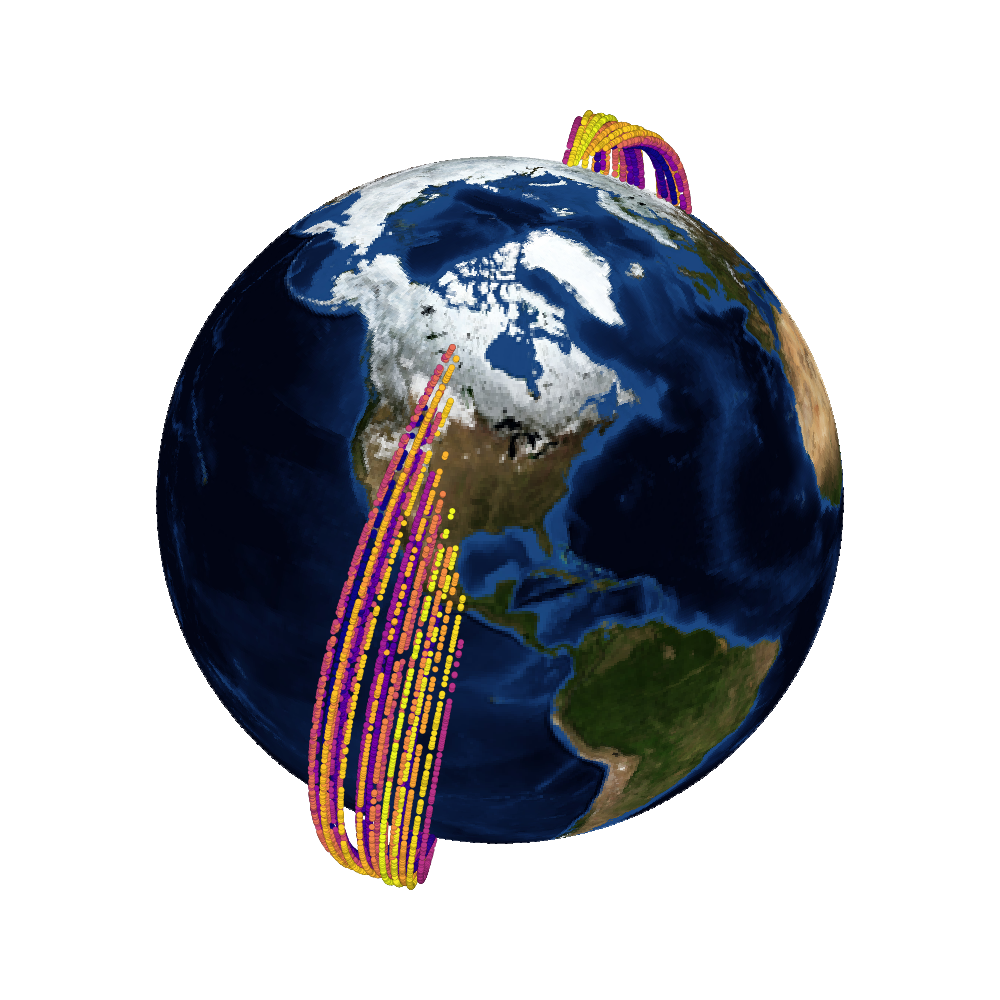}
\includegraphics[width=6cm]{ 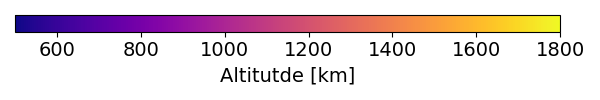}\includegraphics[width=6cm]{ 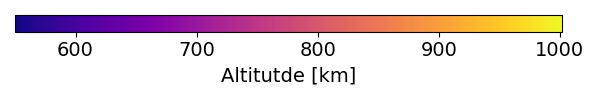}\includegraphics[width=6cm]{ 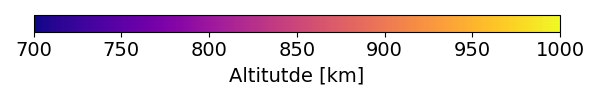}
\caption{Inferred distributions of sunlit satellites using the tight (top) and relaxed (bottom) nodal configurations. 
From left to right: Sunrise (X-ring), SXODC (X-ring plus orbital shell), and Stampede (single ring).  
All distributions are for the December solstice. 
Earth's rotational orientation is arbitrary. 
The colour of the satellite points indicate the relative satellite altitude.
The ``missing'' sections of the Stampede and SXODC X-ring are in shadow.
Sunrise also has a section of its X-ring in shadow, but this is less obvious due to the extended range of its satellites in altitude.
The $30^\circ$ shell in the SXODC will further ensure many satellites go in and out of shadow for some fraction of every orbit, year-round.}
\end{figure}

Figure \ref{fig:sunrise_Earth_map} shows the satellite distributions about Earth for the Sunrise (X-ring), SXODC (X-ring plus orbital shell), and Stampede (single ring) megaconstellations using the tight and relaxed nodal configurations for the Sun-synchronous orbits.
The X-ring configurations for Sunrise and SXODC are apparent, but with SXODC's $30^\circ$ shell creating substantial sky cover.
The tight configuration, as expected, creates well-defined rings, with some warp of the disk to account for the change in Sun-synchronous inclination with orbital altitude (Equation~\ref{eq:sunsynch}). 

The relaxed configuration, in contrast, results in much broader ring systems overall, with crossings now near the poles, but distributed in altitude. 
Recall that the relaxed configuration uses $\pm10^\circ$ uniform node variation centred on the tight configuration.
The images highlight how even seemingly small variances in satellite orbital elements could have a large visual effect, as well as knock-on effects for orbital safety and orbital traffic management.

In the figure, only sunlit satellites are shown, with the Sun-Earth-satellite geometry set for the December solstice. 
Note how the northern sections of one SXODC X-ring and the Stampede singular ring have no illuminated satellites. 
The Sunrise X-ring also has part of a northern section of one of the rings eclipsed, but this is less obvious due to the extended distribution of satellites. 
For the SXODC and Sunrise X-rings, a similar behaviour will be seen during the June solstice, but with eclipses over the southern sections for one of the rings in each X-ring.
In contrast, the single Stampede ring will transition into perpetual sunlight during that time of year. 
This arises from the changing orientation of Earth's shadow (relative to its rotation axis) as the Sun's declination varies.

\begin{figure}\label{fig:tight_sky}
\centering
\hspace*{-1.5cm}
\includegraphics[width=6cm]{ 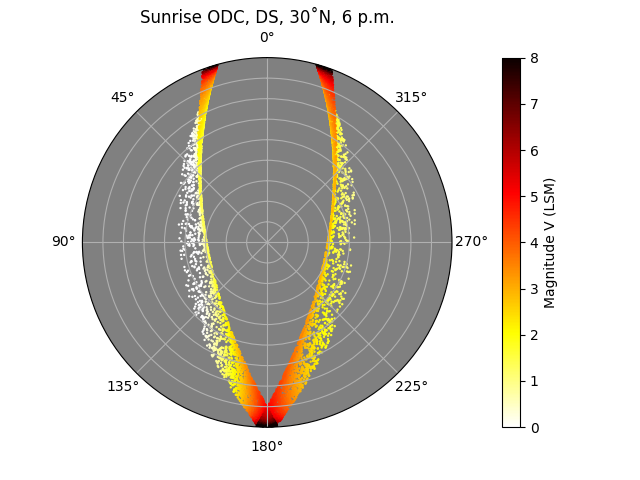}\includegraphics[width=6cm]{ 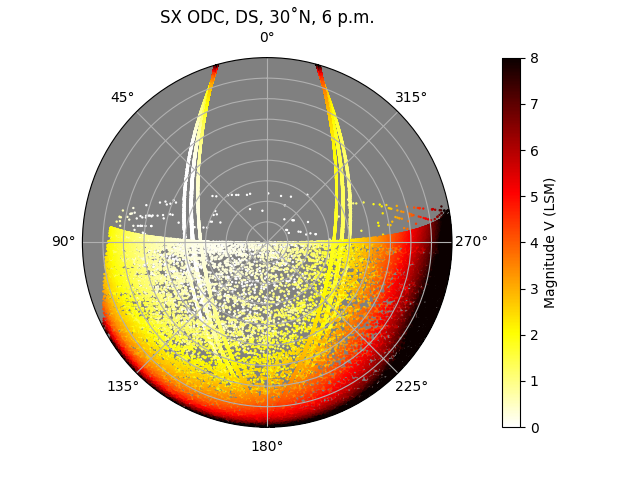}\includegraphics[width=6cm]{ 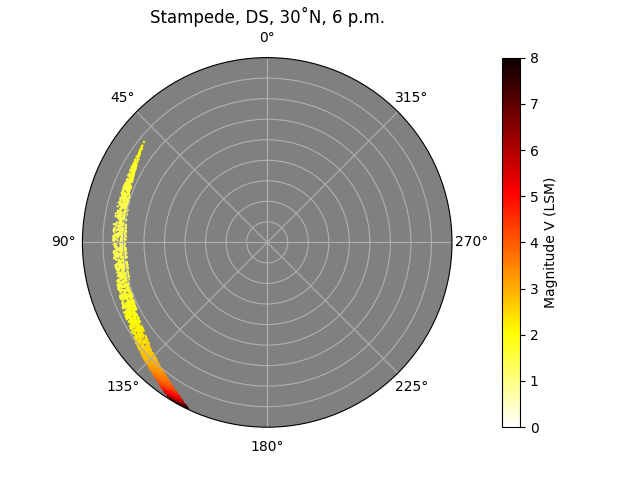}
\hspace*{-1.5cm}
\includegraphics[width=6cm]{ 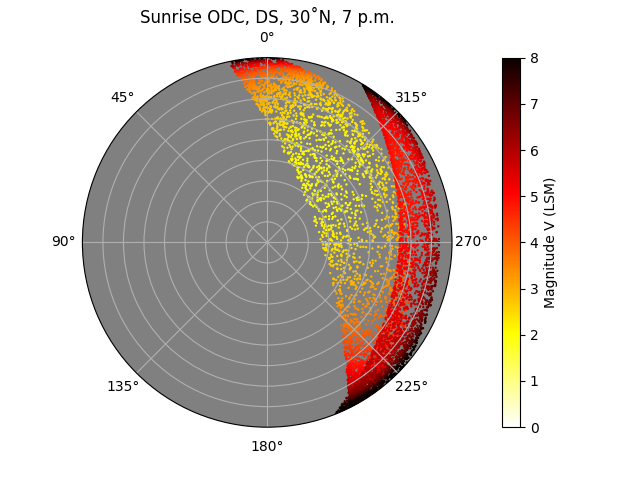}\includegraphics[width=6cm]{ 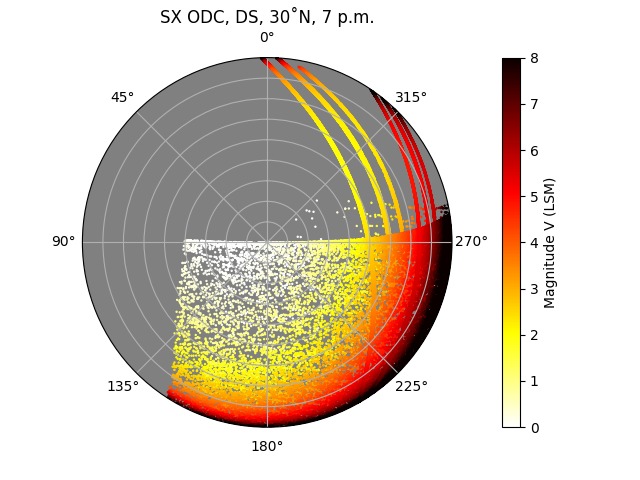}\includegraphics[width=6cm]{ 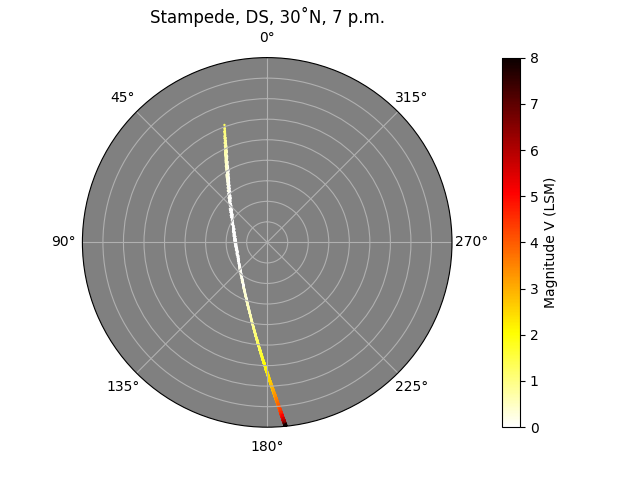}
\caption{Local sky projections using the tight configurations as viewed from $30^\circ$ N on the December solstice for Sunrise (X-ring; left column), SXODC (X-ring plus orbital shell; center column), and Stampede (single ring; right column), at 6~p.m.\ (top row) and 7~p.m.\ local time (bottom row). 
Magnitudes are estimated based on the Lambertian sphere reference.}
\end{figure}

\begin{figure}\label{fig:relax_sky}
\centering
\hspace*{-1.5cm}
\includegraphics[width=6cm]{ 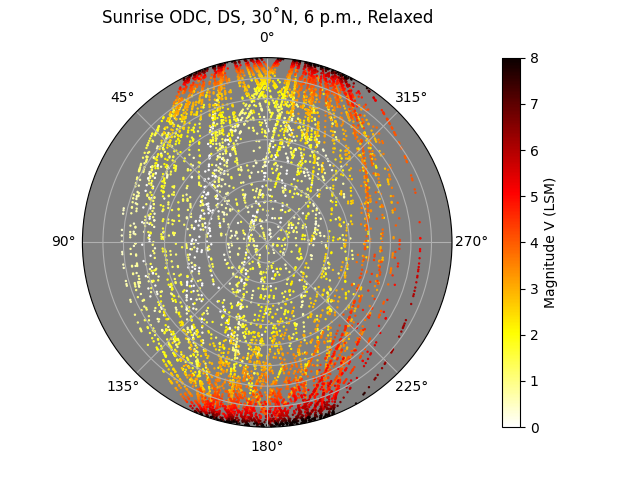}\includegraphics[width=6cm]{ 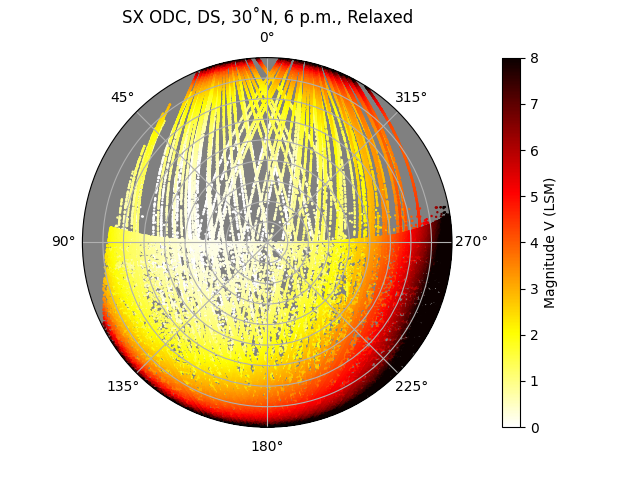}\includegraphics[width=6cm]{ 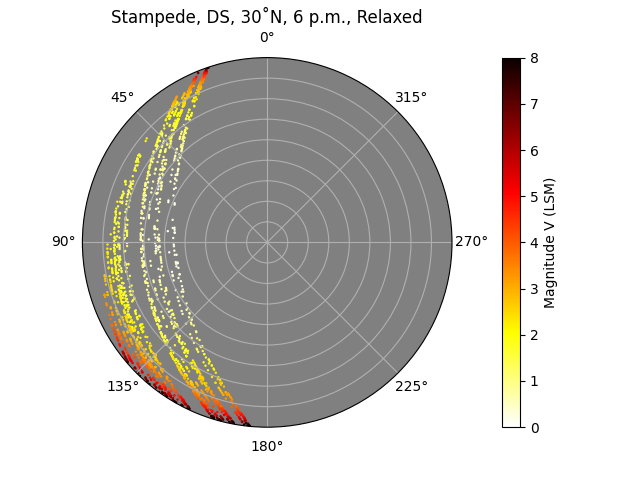}
\hspace*{-1.5cm}
\includegraphics[width=6cm]{ 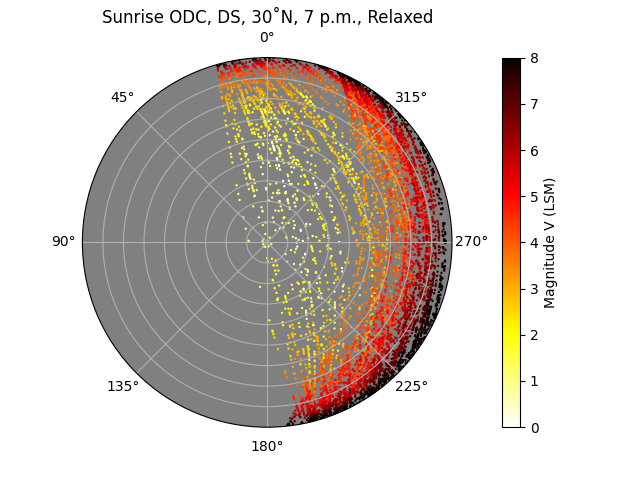}\includegraphics[width=6cm]{ 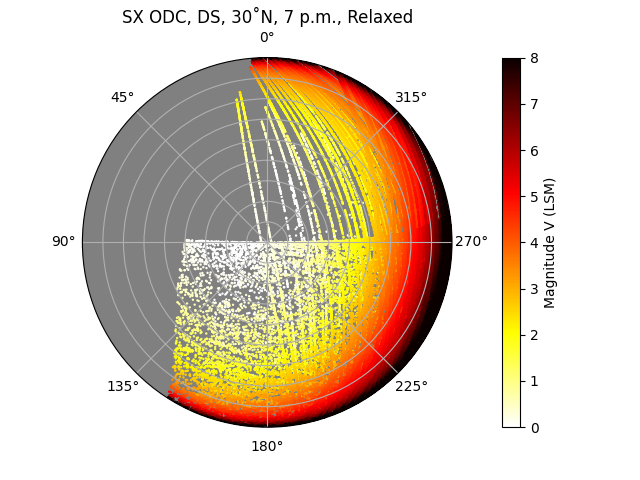}\includegraphics[width=6cm]{ 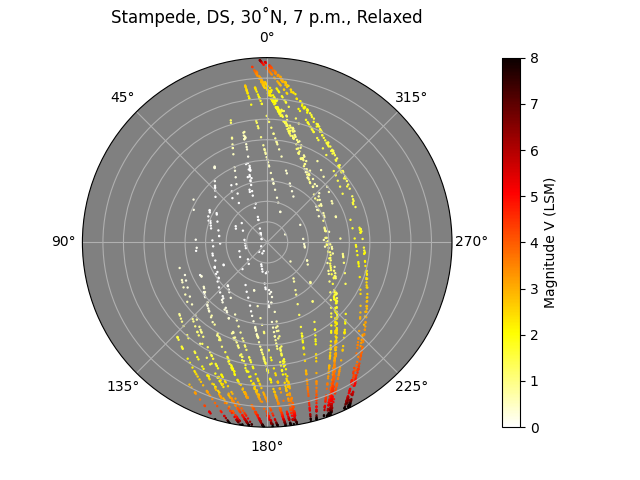}
\caption{Similar to \ref{fig:tight_sky}, but for the relaxed nodal configuration. }
\end{figure}

Figures \ref{fig:tight_sky}-\ref{fig:relax_sky} show the consequences of these configurations in an observer's local sky for 6 p.m. and 7 p.m. local solar time. 
We also provide animated versions of several 
simulations.\footnote{Available at \url{https://www.canfar.net/citation/landing?doi=26.0021}.} 
As discussed in the methods, the predicted satellite magnitudes are based on the LSM and there can be considerable variation (dimmer and brighter) in satellite brightness in practice.

Overall, the ODCs are expected to be very bright and create noticeable sky structures. 
The tight nodal configurations will have clear ring segments that wrap through the sky, while the relaxed configurations will instead create widespread sky cover. 
At low to mid-latitudes, the SXODC and Sunrise impacts are  largest around 6 p.m./6 a.m. and evolve across the sky quickly. 
The $30^\circ$ SXODC shell further causes some sections of the sky to be densely covered by very bright satellites. 
When paired with the visible Sun-synchronous rings, the SXODC has an appearance of overalls rotating across the sky. 

The Stampede configuration requires a special mention. 
Its node is not exactly at a LTAN of 6 p.m., but is instead delayed by 48 minutes. 
This has two effects. 
One is that the shadowed portion of the ring is moved westward.
The other is that the shift in LTAN causes the ring to be highest in the sky later in the evening (with the 7 p.m. image showing a brighter and higher ring arc than at 6 p.m.). 
The flip side is that ring rise is delayed more into twilight, although satellites can still be high in the sky and bright at 6 a.m. in a dark, winter sky. 

\begin{figure}\label{fig:sxodc_year}
\centering
\hspace*{-1.5cm}
\includegraphics[width=0.49\textwidth]{ 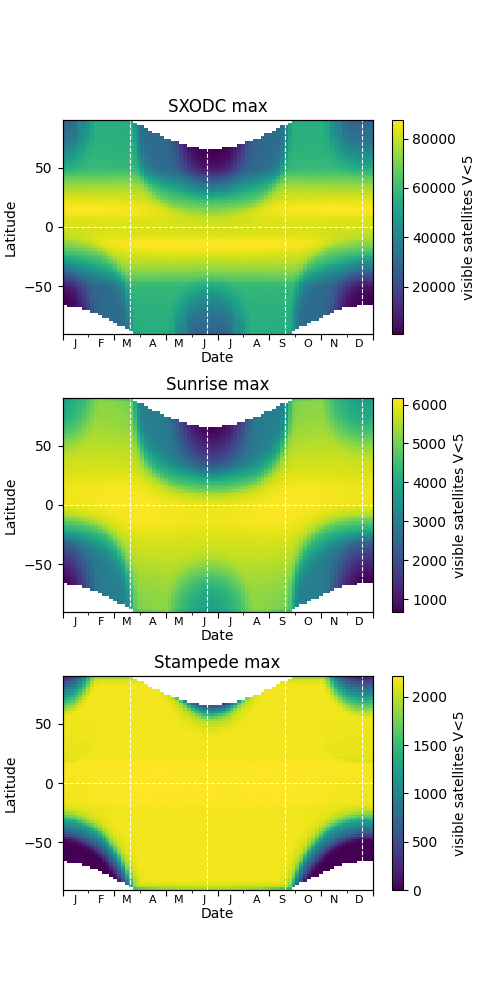}
\includegraphics[width=0.49\textwidth]{ 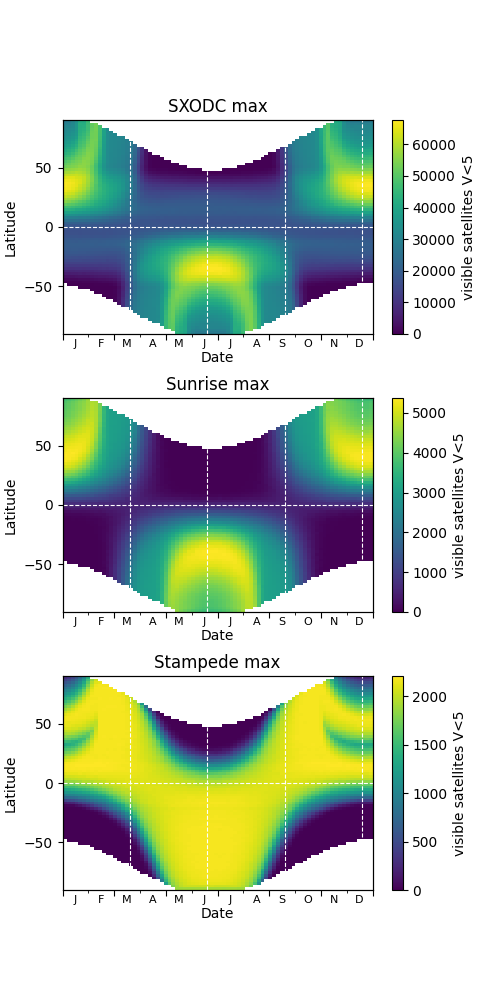}
\caption{The maximum number of visible satellites brighter than V $<$ 5 during the night for different latitudes and dates. 
Only satellites with altitudes $10^\circ$ above the horizon are included.
The top plots shows SXODC, the middle ones Sunrise, the bottom ones the Stampede constellation.  
The left plots show the maximum number of visible satellites between sunset and sunrise.
The right plot shows the maximum number of visible satellites between the end of astronomical twilight in the evening to the beginning of astronomical twilight in the morning.
The polar regions are removed when they do not experience darkness.
}
\end{figure}

\begin{figure}\label{fig:odc_timeofday}
\centering
\hspace*{-1cm}
\includegraphics[width=0.32\textwidth]{ 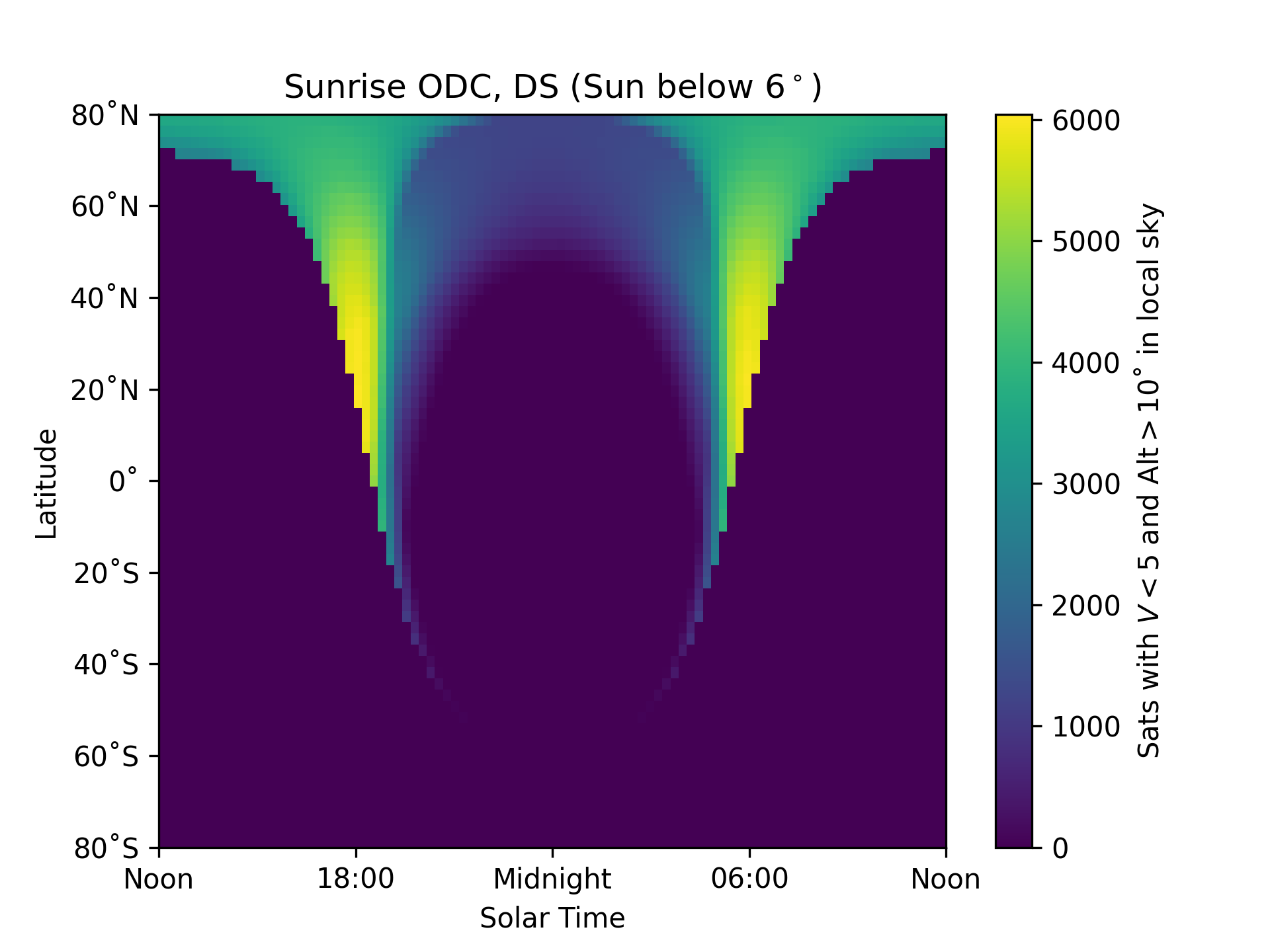}\includegraphics[width=0.32\textwidth]{ 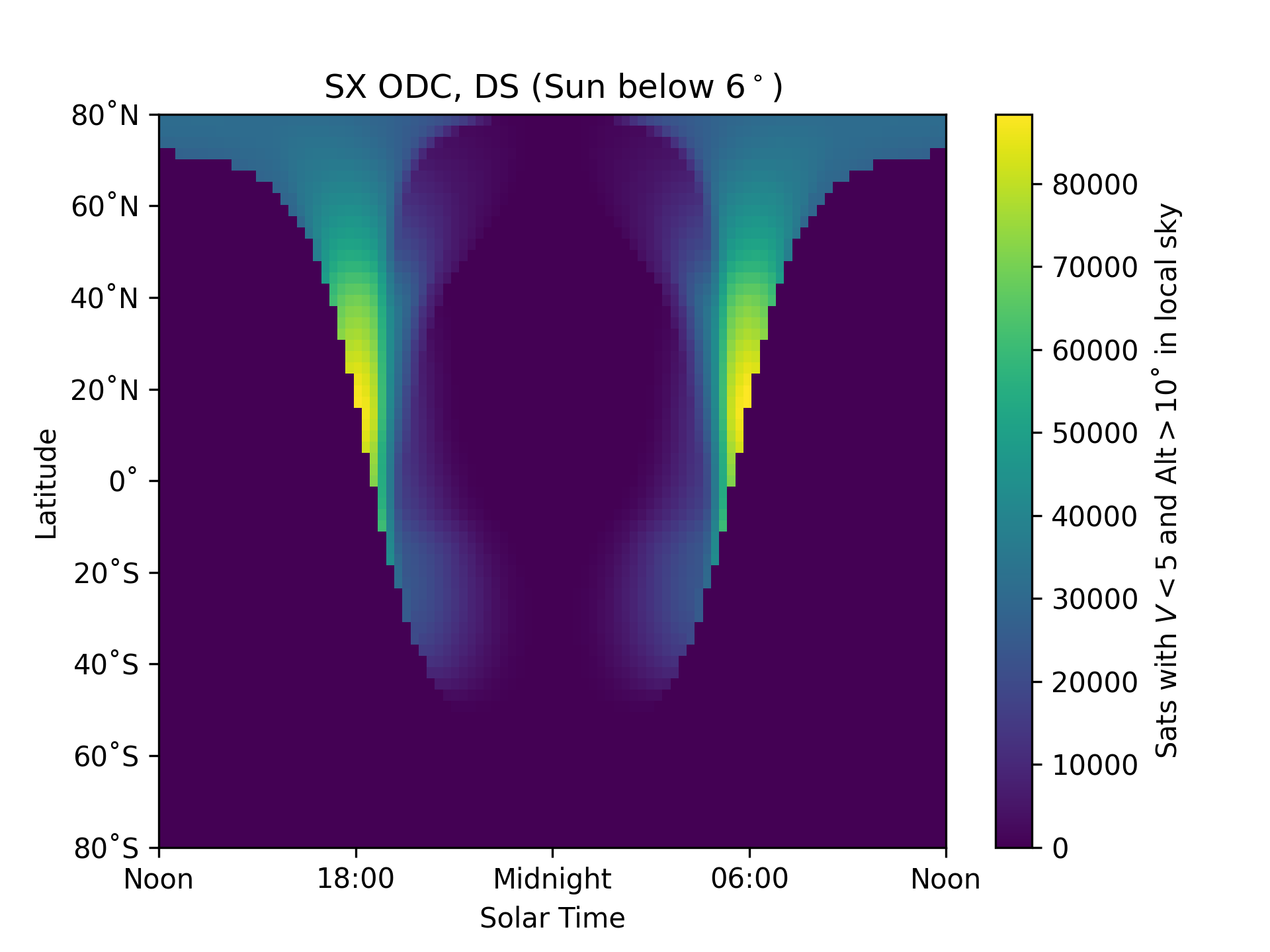}\includegraphics[width=0.32\textwidth]{ 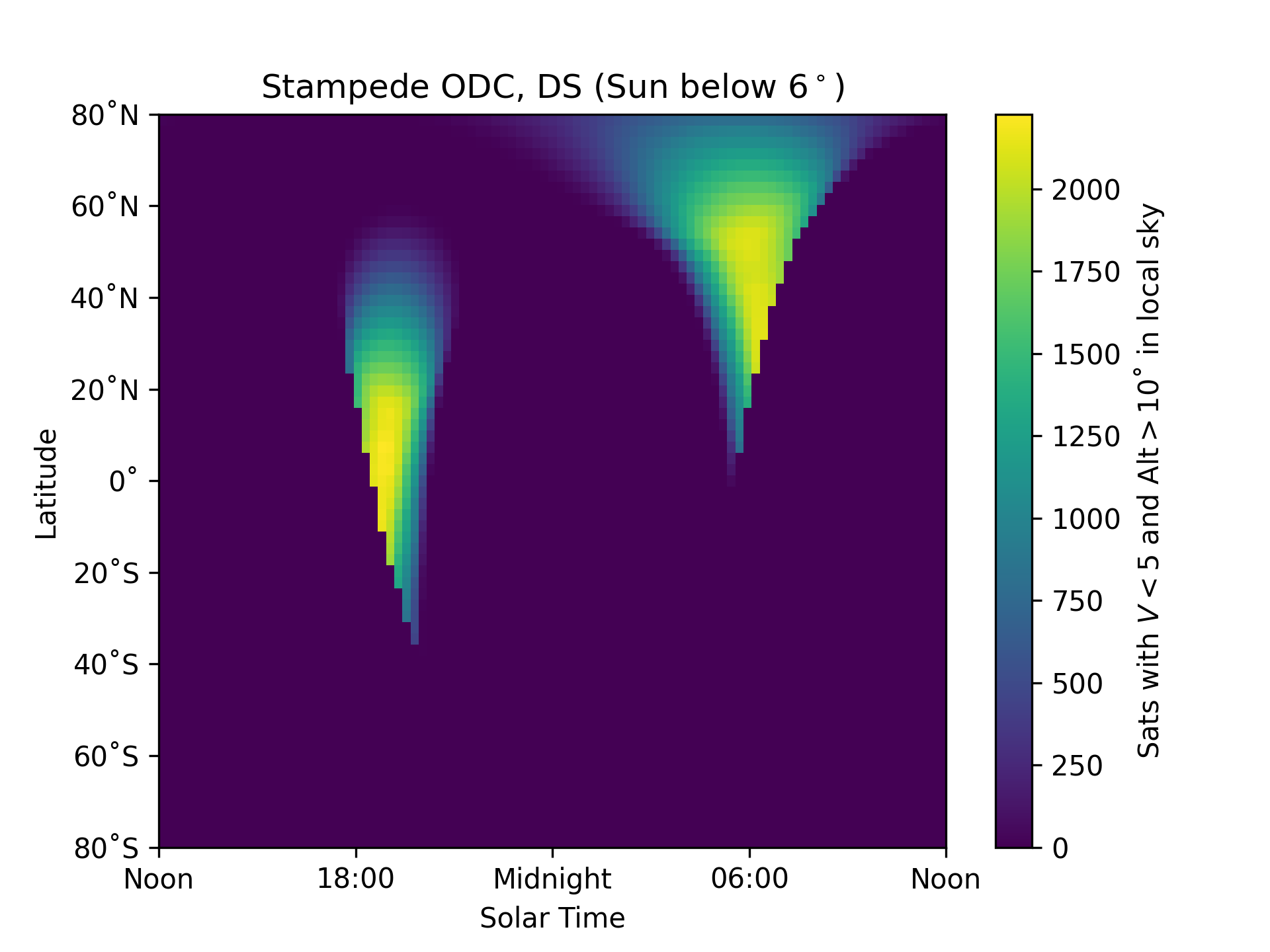}
\hspace*{-1cm}
\includegraphics[width=0.32\textwidth]{ 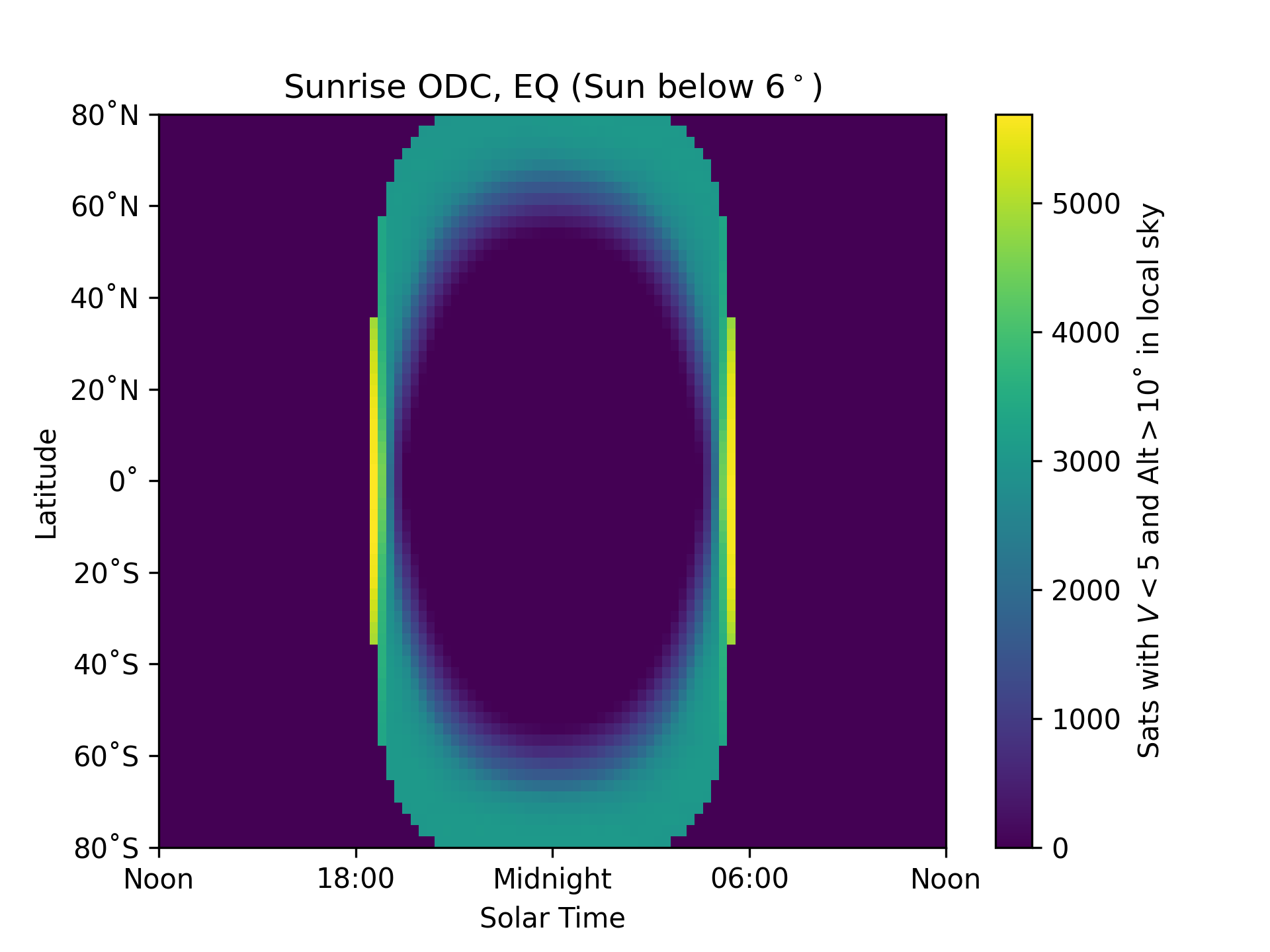}\includegraphics[width=0.32\textwidth]{ 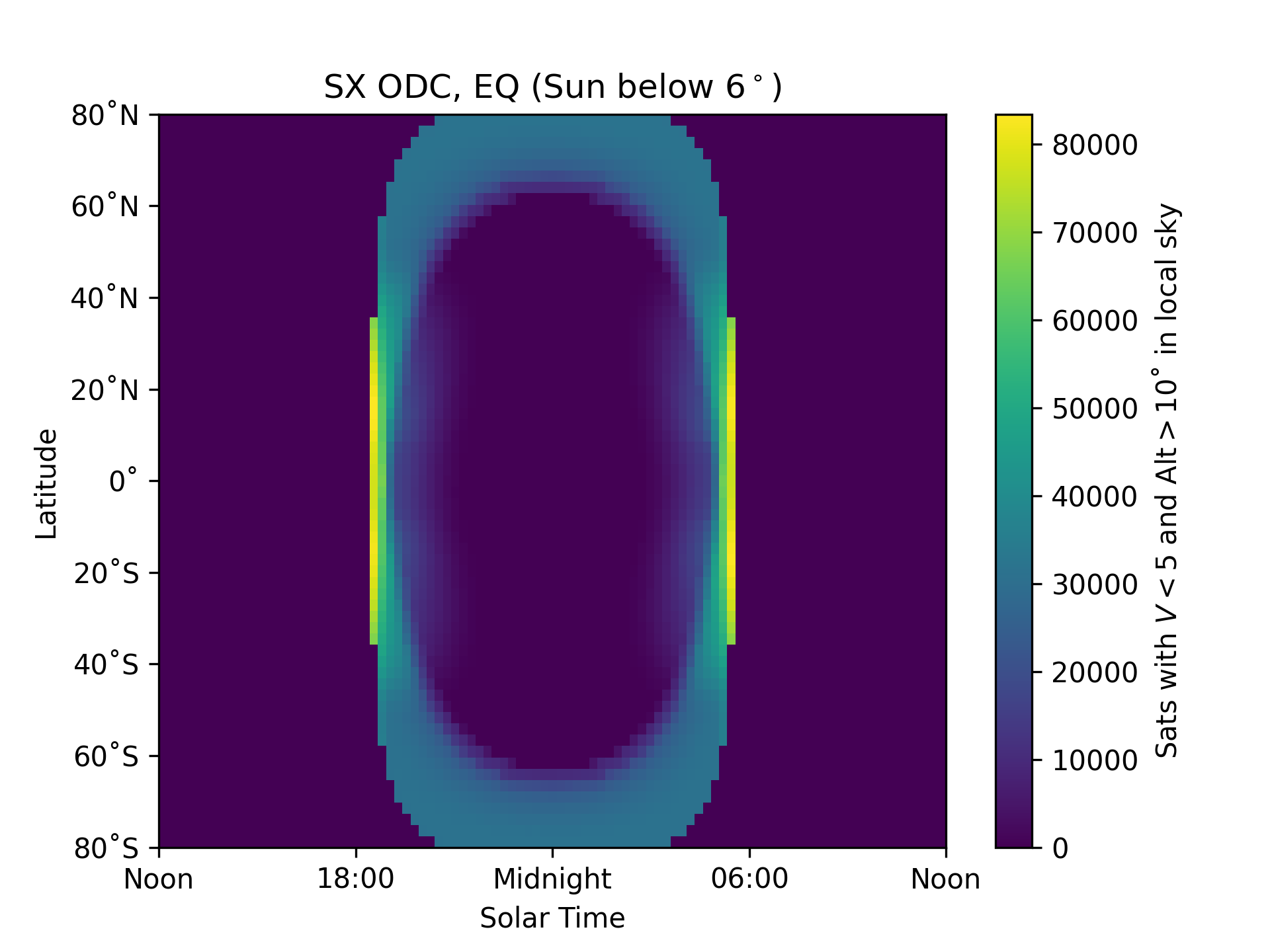}\includegraphics[width=0.32\textwidth]{ 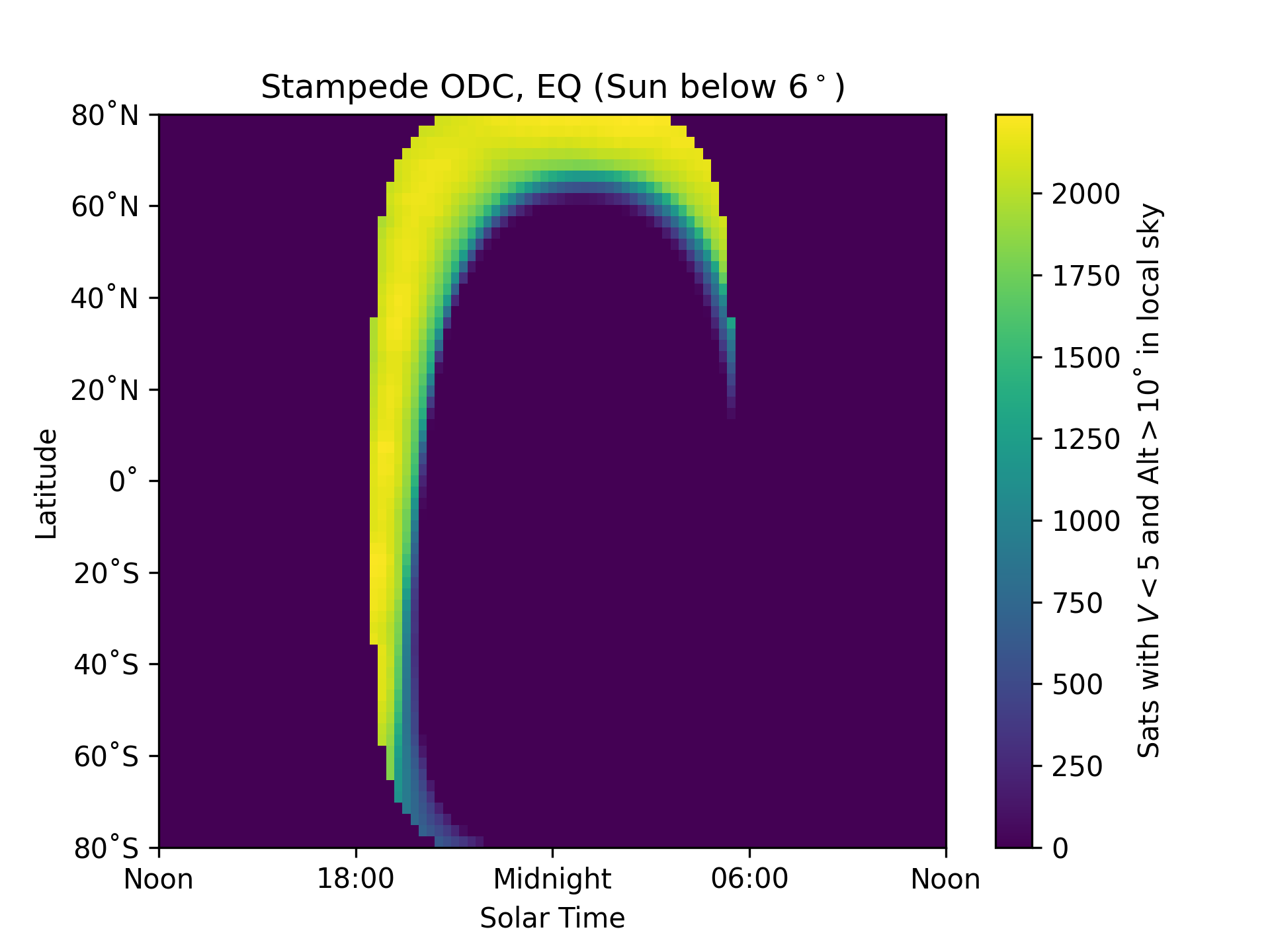}
\caption{The number of visible satellites brighter than V $<$ 5 throughout the night and for different latitudes.
Satellites are only shown when the Sun is $6^\circ$ or more below the horizon (i.e., darker than civil twilight).
Only satellites with altitudes $10^\circ$ above the horizon are included in the counts.  
The top shows the results for Sunrise, SXODC, and Stampede for the December solstice, while the bottom is for the equinox.
The strong asymmetry in the Stampede panels is due to the 48-minute offset in LTAN. 
This same asymmetry is seen in Figure \ref{fig:sxodc_year}.
}
\end{figure}

Because these ODCs are enormous compared with current LEO satellites, the LSM predicts that they could be extremely bright when sunlit.  
Table \ref{tbl:brightness} and Figure~\ref{fig:sxodc_year} demonstrate that at certain times across a wide range of latitudes the proposed ODC megaconstellations could produce more visible satellites than naked-eye stars, including more bright satellites than bright stars. 
Specifically, Table \ref{tbl:brightness} presents the count of satellites above different magnitude cutoffs for a local observer at 6 p.m. for SXODC and Sunrise and at 7 p.m. for Stampede at a latitude of $30^\circ$ (north or south) during local winter.
While most of the reported numbers are for the entire visible sky, one column shows the number of visible satellites above $10^\circ$ altitude for $V<5$.
For comparison, the last column shows real star counts based on a random local sky snapshot of the Yale Bright Star Catalogue \citep{bright_star_catalogue,vizier}, with the same extinction function applied as done here for satellites. 
 Using the extinction function gives a better sense of what someone could count with excellent eye sight and in an exceptionally dark location.
The real star counts emphasize that there are typically about 500 stars with $V<5$ that an observer can see in their local night sky with the unaided eye, accounting for atmospheric extinction. 
This magnitude cutoff roughly corresponds to what one could see in a moderately light-polluted sky.

Figure~\ref{fig:sxodc_year} highlights how dramatically the night sky could change with so many large satellites.  
It shows the number of $V < 5$ SXODC satellites above $10^\circ$ altitude in an observer's local sky for different latitudes and different times of year, at both midnight and sunset (if applicable). 
At sunset in particular, there could be tens of thousands of bright satellites visible from nearly every position on Earth in every season.  
At midnight, the polar regions could continue to experience tens of thousands of bright satellites (this is discussed further in Section~\ref{sec:disc} below).

The structure seen in Figure \ref{fig:sxodc_year} is emphasized by the 24-hour plots in Figure \ref{fig:odc_timeofday}, which show the number of visible satellites with $V<5$ as a function of observer latitude and time of day. 
Only satellites that are $10^\circ$ above the horizon are included in the counts. 
Moreover, satellite numbers are only shown if the Sun is $6^\circ$ degrees or more below the observer's local horizon (i.e., between the end and start of civil twilight).
Sunrise and SXODC are symmetric, while Stampede shows a strong asymmetry due to the offset of its LTAN. 
That offset in turn places the north-western section of the ring in shadow during the December solstice in addition to creating a timing difference. 
Because the Stampede ring is fully illuminated during the June solstice, the asymmetry produces a different signature for the June solstice for observers in the southern hemisphere (not shown).
The asymmetry in the December solstice plot for Stampede ultimately produces the double peak in satellite counts near the December solstice as a function of latitude in the Stampede panel of Figure \ref{fig:sxodc_year}.

When ODC rings are visible at night, they could become more prominent than the Milky Way, with potentially many far brighter satellites than bright stars, should mitigations fail.

%NOTE: Values computed using numpy.random.seed(314)
\begin{table}
    \centering
        \caption{Brightness distributions for tight (T) and relaxed (R) configurations\\At Latitude $30^\circ$ N}
        \label{tbl:brightness}
        \begin{tabular}{l|c|cccccc}\hline
                      & Above $10^\circ$ & \multicolumn{6}{c}{Full Local Sky}\\
    Megaconstellation & $V<5$ & $V<0$ & $V<3$ & $V<4$ & $V<5$ & $V<6$ & $V<7$ \\\hline
       Sunrise T (6 p.m.) & 6000 & 280 & 4400 & 6100 & 7100 & 7700 & 8000\\
       Sunrise R (6 p.m.)& 5700 & 210 & 4100 & 5600 & 6800 & 7500 & 7800\\
       SXODC T (6 p.m.)& 79000 & 5100 & 66000 & 85000 & 98000 & 110000 & 110000\\
       SXODC R (6 p.m.)& 75000 & 3900 & 59000 & 78000 & 93000 & 100000 & 110000\\
       Stampede T (7 p.m.) & 1500 & 320 & 1300 & 1500 & 1700 & 1800 & 1800\\
       Stampede R (7 p.m.) & 1500 & 93 & 1200 & 1500 & 1700 & 1900 & 1900\\\hline
       %Sunrise T (6 p.m.) & 5986 & 283 & 4435 & 6095 & 7084 & 7669 & 7996\\
       %Sunrise R (6 p.m.)& 5745 & 208 & 4093 & 5613 & 6844 & 7482 & 7849\\
       %SXODC T (6 p.m.)& 79060 & 5103 & 66443 & 85014 & 98083 & 106543 & 112586\\
       %SXODC R (6 p.m.)& 75407 & 3867 & 59366 & 78334 & 92928 & 102337 & 108906\\
       %Stampede T (7 p.m.) & 1474 & 316 & 1271 & 1500 & 1677 & 1775 & 1831\\
       %Stampede R (7 p.m.) & 1525 & 93 & 1230 & 1534 & 1746 & 1867 & 1940\\\hline
       Bright Stars & 470 & $\lesssim1$ & 56 & 180 & 500 & 1700 & 4000 \\\hline
       \end{tabular}

\end{table}

\section{Discussion: How much will our night sky change?} \label{sec:disc}

ODCs are designed to be sunlit as much as practicable.
This pushes their configurations to have most if not all of the proposed orbits above altitudes of 600 km.
Moreover, to attempt to achieve perpetual illumination, the rings must have terminator-polar orbits -- although as discussed in section \ref{sec:shade}, this alone is insufficient, as the rings will have eclipse seasons unless they are at very high altitudes ($\gtrsim$1400~km).
At mid-latitudes, such rings naturally place the satellites well above the horizon around local 6 p.m. and 6 a.m., quickly evolving across the sky over the next two hours.
During the summer, again at mid to low-latitudes, the rings will be highest in the sky during daylight, avoiding the worst night-sky impacts.
However, during winter, the satellites are above the horizon during late twilight hours and into some portions of the night.
For example, during June and July, local 6 p.m. in La Serena, Chile, (about $30^\circ$ S) will roughly correspond to the start of astronomical twilight, creating the conditions for bright rings in near-night conditions. 
For Stampede, which has a shifted LTAN, the ring will be highest closer to 7 p.m.

At mid-high latitudes (near $50^\circ$ N and S) in the wintertime, sections of the rings will be visible rising and then moving across the sky as soon as it is dark. 
And from high latitudes (poleward of $60^\circ$ N or S), a portion of the ring will be visible during the entire long winter night. 
Even during the equinoxes, the data centres could become among the first ``stars'' visible each night.
Radio astronomy will be affected regardless the time of year within several hours of observatory terminator crossing. 

The $30^\circ$ orbital shell proposed in the SXODC would have a sky impact similar to what is seen with existing megaconstellations, just much more severe due to higher on-sky densities and much larger satellites.
The $30^\circ$ inclination also places the most severe impacts on mid and low-latitude locations.
For example, at a latitude of $50^\circ$ N/S, the $30^\circ$ shell would have a highest altitude in the sky of about $10^\circ$, reducing the impact of this particular component.
However, as shown in Figure \ref{fig:tight_sky}-\ref{fig:relax_sky}, the shell could cover large portions of the sky as seen from temperate to equatorial latitudes.
This could include, depending on the actual on-sky brightness, having more visible satellites than stars shining throughout twilight and into portions of night.

\begin{figure}\label{fig:polar_sky}
\centering
\hspace*{-1.5cm}
\includegraphics[width=6cm]{ 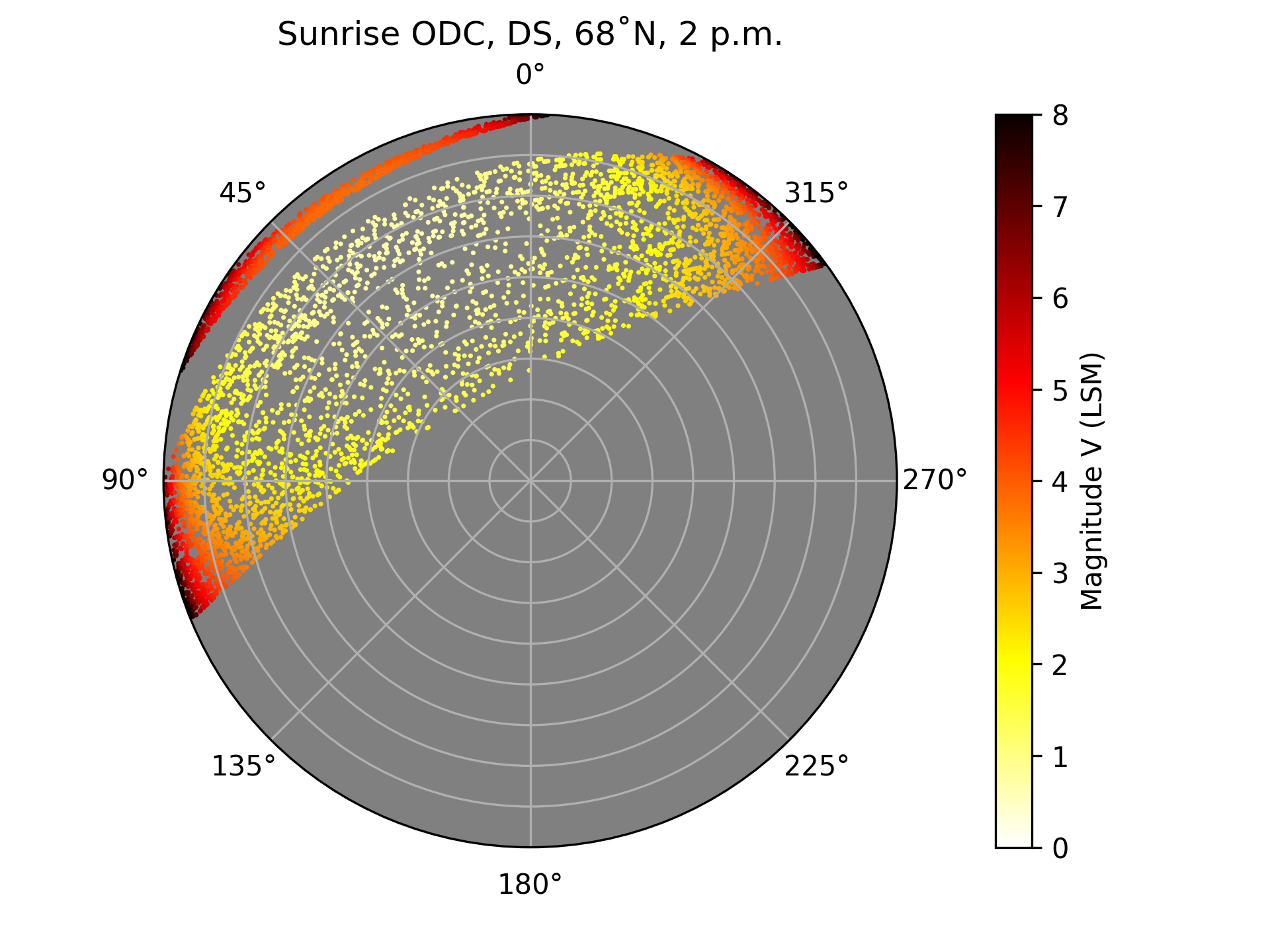}\includegraphics[width=6cm]{ 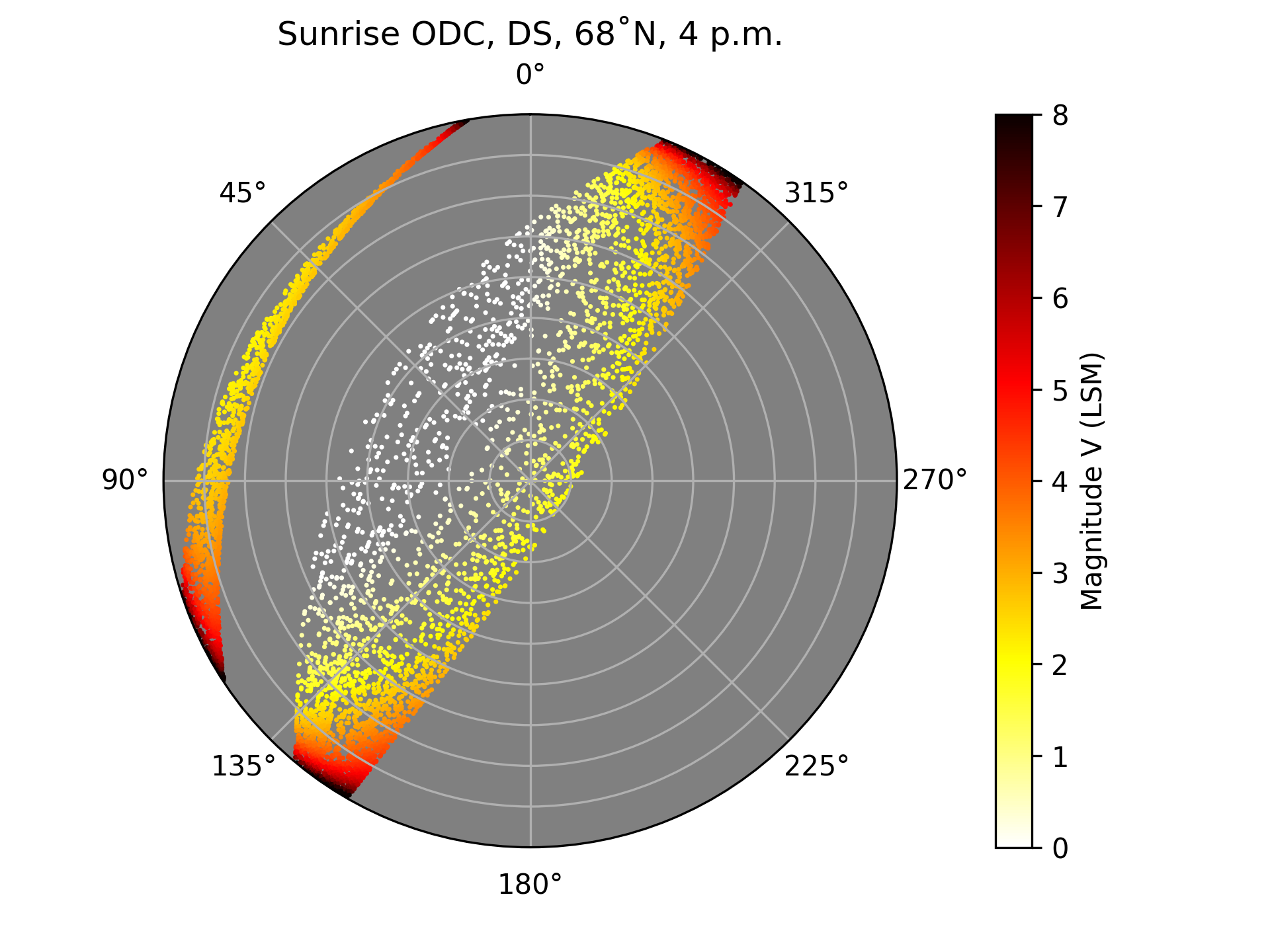}\includegraphics[width=6cm]{ 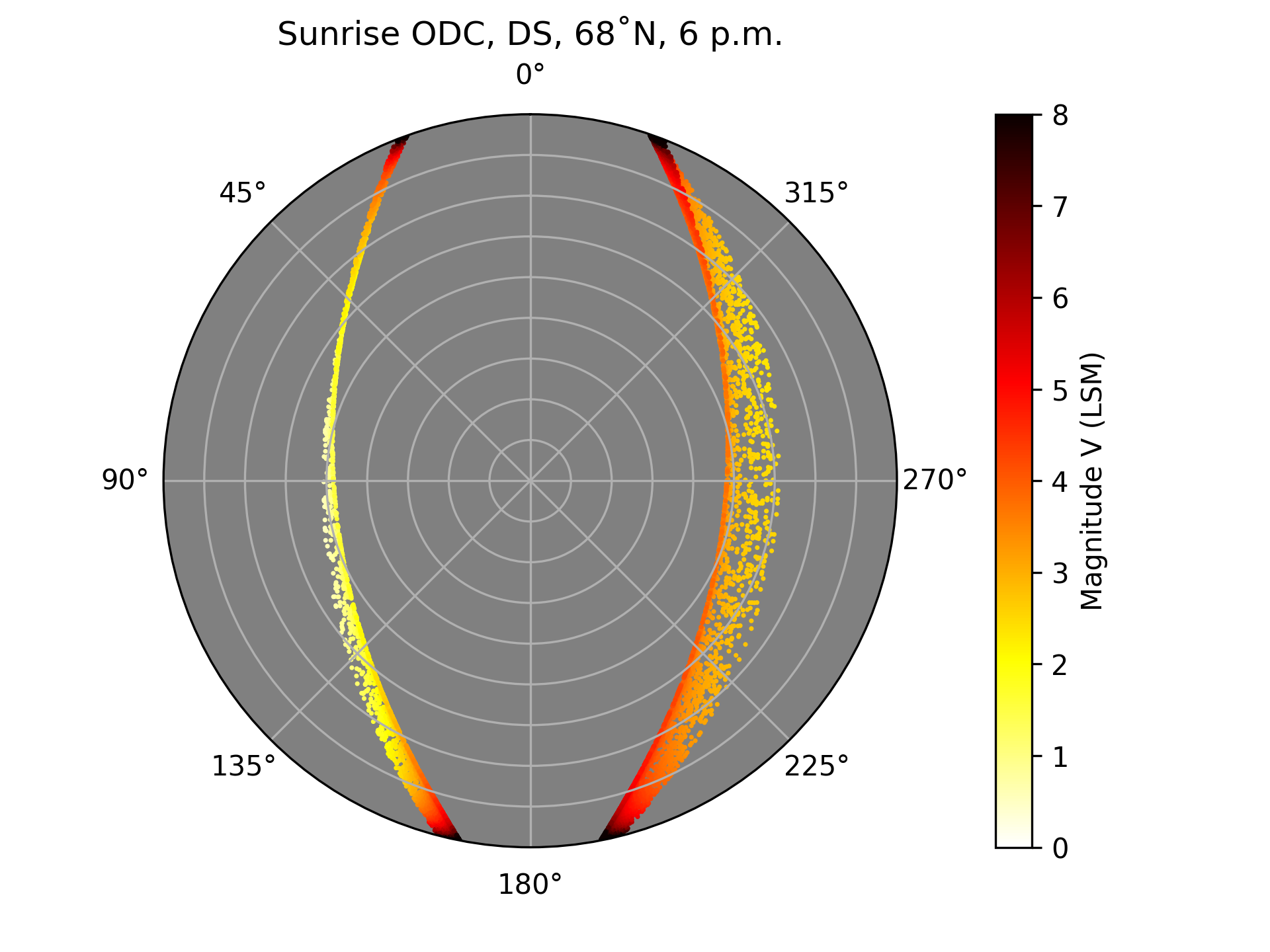}
\hspace*{-1.5cm}
\includegraphics[width=6cm]{ 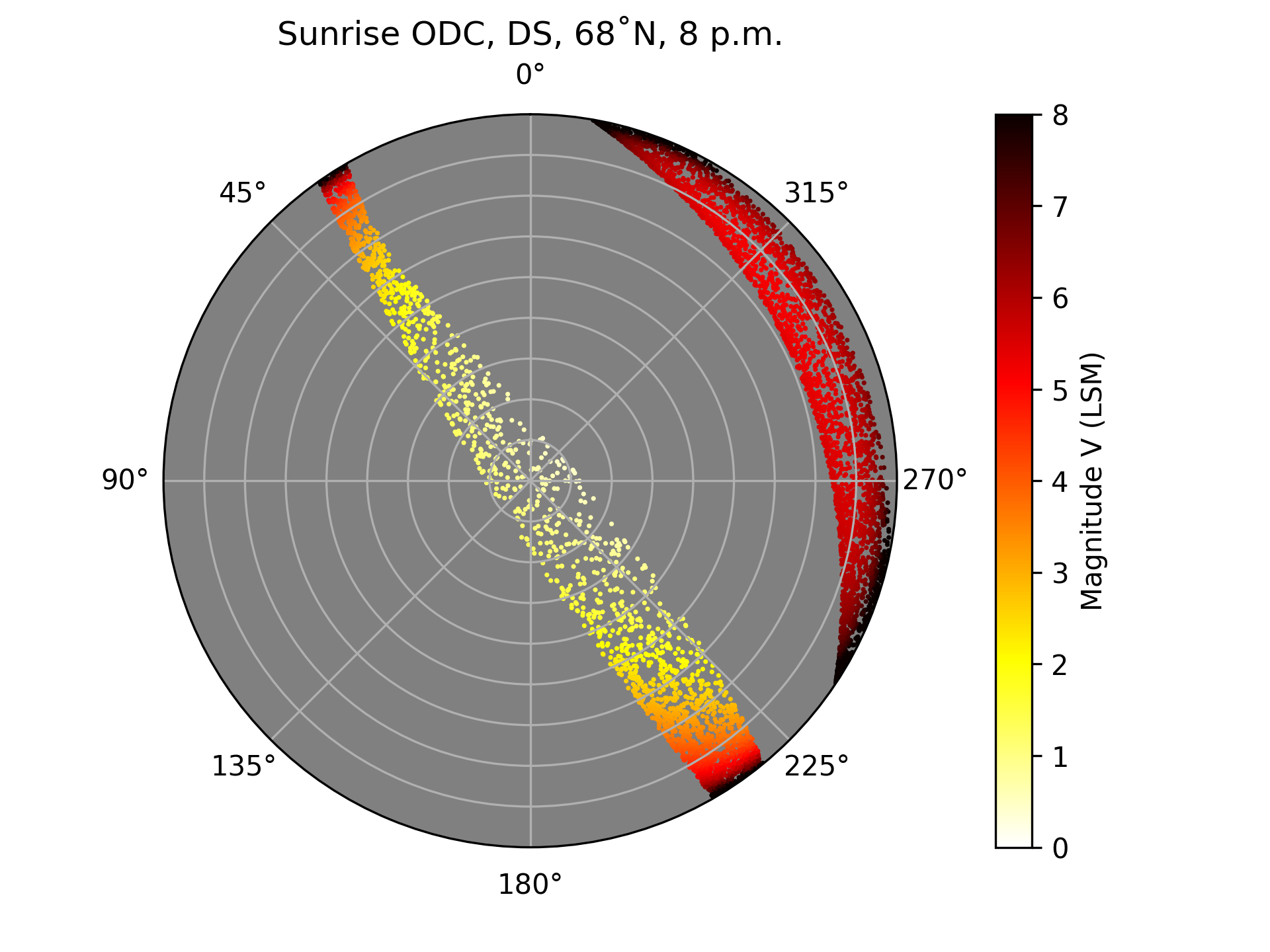}\includegraphics[width=6cm]{ 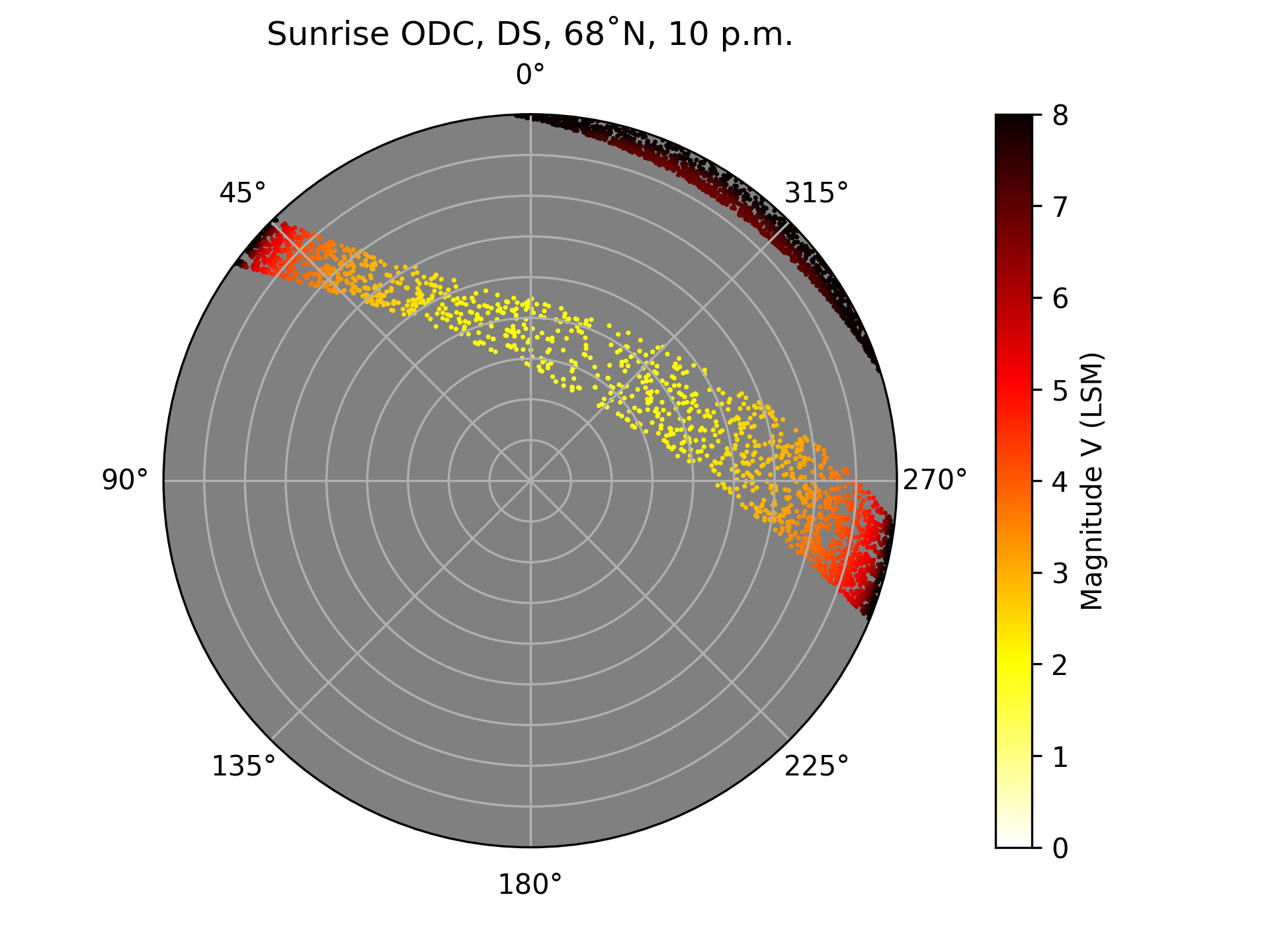}\includegraphics[width=6cm]{ 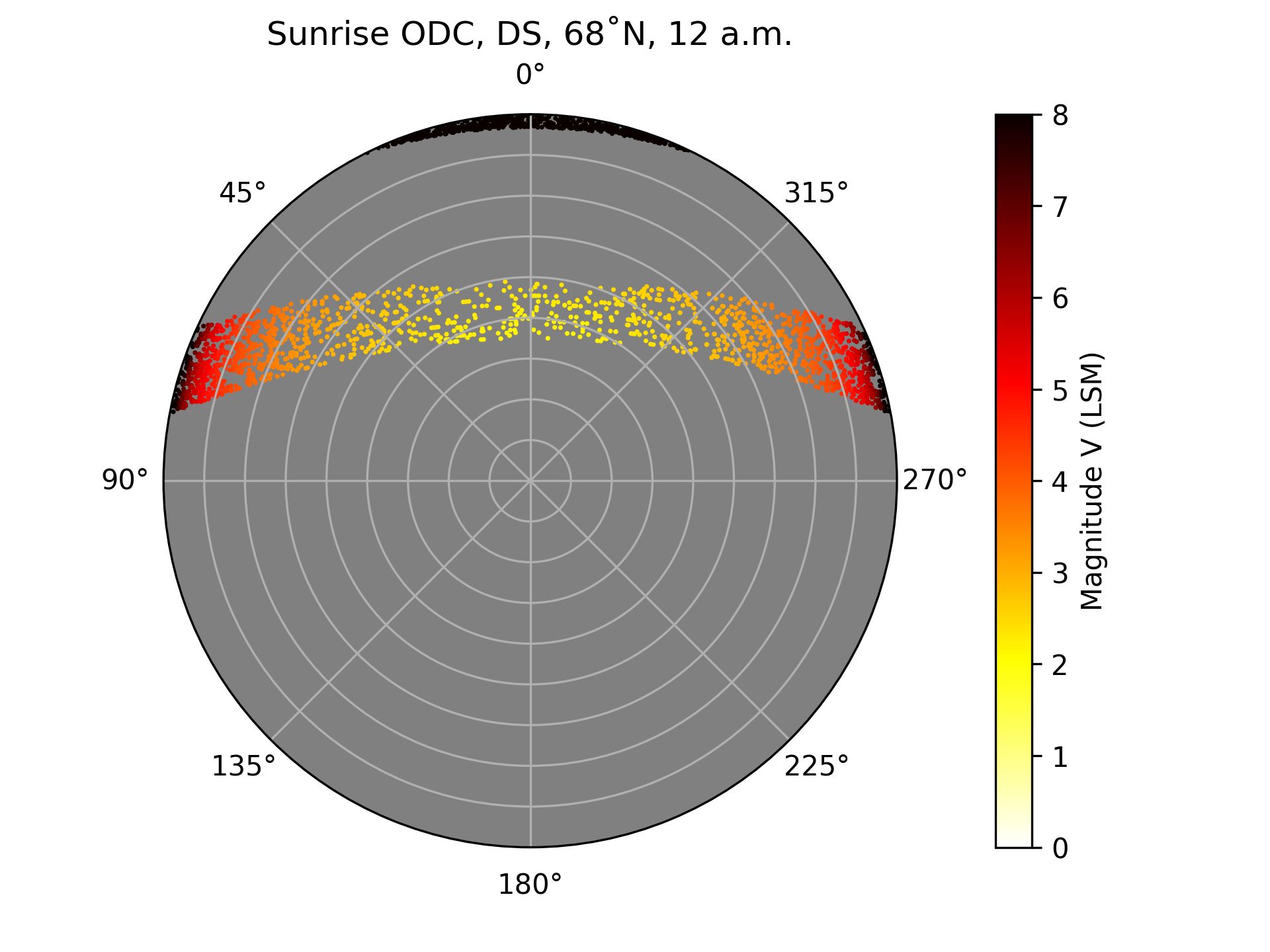}
\caption{Sunrise ODC seen at $68^\circ$N (such as Inuvik, NWT, Canada) on the December solstice, when the observer is in twilight or night for 24 hr.
A similar picture would be seen at Antarctic research stations during the June solstice, although many of the stations will be at even higher latitudes.}
\end{figure}

The impacts on Earth's polar skies requires additional discussion.
Polar-terminator rings will always have arcs above the horizon within the Arctic and Antarctic Circles.
As those arcs change orientation throughout the day and night, the thickness and the brightness of the rings (even tight nodal configurations) will change as the observer rotates through different viewing angles (see Fig.~\ref{fig:polar_sky}).
In the northern hemisphere, communities in, for example, the Canadian North will potentially see satellite rings for 24 hours during winter and also throughout night close to the equinoxes. 
In the southern hemisphere, the situation is similar, but with potential impacts on Antarctic-based science, whether optical, infrared, or radio experiments. 

Radio, infrared, and possibly submillimetre astronomy will not have any reprieve from Earth's shadow. 
While not modelled here, we note that the satellites will need to shed large amounts of heat using large onboard radiators.
The ODCs and their radiators could be of concern for infrared and submillimetre observations.
Unintended electromagnetic radiation (UEMR) and possibly reflections of terrestrial signals could further cause substantial interference with radio observatories. 

For tight-ring configurations, the locations of ODC arcs will be constrained and predictable, though scattered light could still be an issue, as discussed below.
Moreover, different projections of the rings can still cause large satellite sky coverage.
Nonetheless, this setup offers the possibility of at least some pointing avoidance for astronomical observations.
In contrast, the relaxed versions explored here ($\pm 10^\circ$ for the node) essentially blanket the sky with satellites when the rings are above the horizon, as shown in Figures \ref{fig:tight_sky}-\ref{fig:relax_sky}, making pointing avoidance impractical, in general.
Moreover, multiple operators with multiple rings will further interfere with pointing avoidance, whether by optical or radio observatories.
Radio pointing avoidance could be further complicated by telescope sidelobes.

Recent modelling by \citet{Hainaut2026} has shown that large numbers of bright satellites will significantly affect an observer's sky brightness, with effects distributed widely across the sky.
While such modelling was not specifically for ODC rings, it nonetheless shows that a high density of on-sky bright satellites will lead to atmospheric scattered light effects that distribute the ODC impacts across a broader portion of the visible sky than just the projected ring. 
As such, night sky losses are more than the satellites seen individually.

Our models assume all satellites are in their operating orbits and are no worse than having a Lambertian sphere in orbit. 
This sidesteps several important considerations.
For one, the proposed satellite numbers will necessitate a large number of satellites undergoing raising and lowering operations.
The public filings of ODC megaconstellations  estimate the lifetime of each ODC in their ``Schedule S'' to be about five years, but the time needed for raising and lowering is less clear.
Absent additional information, we can use existing megaconstellation behaviour as a benchmark. 
For Starlink, it typically takes months to raise and lower orbits at the beginning and end of their operating lifetimes, respectively \citep{mcdowell_website}.
Thus, if we assume that (1) raising and lowering each take two months and (2) the typical lifetime of an ODC is 5 years, then there will be about 7\% of ODCs away from their operational altitude.
Depending on the raising, lowering, and deployment/unfolding practices, this could result in substantial variation in ODC brightness.

Another concern is the loss of control of a satellite. 
This could lead to tumbling, and with such large surface areas as proposed for ODCs, the impacts of glints should be expected to have serious ramifications for the night sky (in addition to space safety).
Based on \cite{mcdowell_website}, the failure rate of Starlink satellites while on orbit is about 1\%. 
As with raising and lowering, this could imply a major increase in the frequency of very bright glints. 
An on-orbit explosion of an ODC would be serious space safety hazard, and would further create conditions for a high frequency of glints and transients in astronomical data.

Finally, our assumptions of the LSM might significantly underestimate the brightness of satellites in certain cases.
Substructures on the satellites, such as those associated with solar panels, radiators, and the chassis, could lead to regular flaring if not accounted for beforehand, such as what occurred in the Iridium constellation (i.e., Iridium flares). 

\section{Conclusions: How do we keep dark sky access?}

Overall, our calculations show that there is substantial potential for severe night sky impacts should there not be extensive dimming efforts by operators, along with near-zero failure tolerances.
Mitigations may include building brightness reductions into satellite designs and maintaining a high compliance of operational control of the satellites. 
Should ODC megaconstellations be pursued, states and their regulators will need to recognize that (1) clear commitments to brightness mitigations are necessary; (2) impacts will be cumulative with other operators, which can be supervised by other states and regulators; and (3) compliance with operational safety and brightness mitigations will need to be stricter than pursued under past situations. 
The latter point is critical for space safety, but also dark and quiet skies. 

The need for oversight regarding satellite brightness is emphasized by our models, which show in Table \ref{tbl:brightness} the potential severity of on-sky impacts. 
If satellite brightness is not significantly reduced below equivalent Lambertian sphere reflections (by orders of magnitude), satellite rings will become the most prominent night sky features near twilight, especially close to a hemisphere's winter solstice. 
Should the full SXODC be launched, then there could be about 100 times more visible satellites in the night sky than visible stars.

\textit{This new, anthropogenic sky feature would even be easily visible in a light-polluted city: the natural patterns of the star-filled night sky that have been a source of wonder and knowledge for all of human history will have been erased by urban light pollution and then replaced by human-made artificial patterns.}
While it is tempting to hope that systems such as the SXODC have dramatically overfiled orbits, the growing number of proposed ODC megaconstellations means the cumulative impact could still reach extreme levels, if allowed.  

In relation to the IAU recommendations, ODC Sun-synchronous rings are incompatible with the altitude threshold \citep{Boley2025}. 
Given their extensive numbers, as proposed, their impact should be expected to be substantially greater than current Sun-synchronous systems.
As noted above, ODCs will need to reduce their brightness below the equivalent Lambertian sphere by orders of magnitude. 
However, such reductions will inherently not address impacts on orbital observatories, and could in fact make them worse. 
Without strong mandates, incentives, and compliance measures, there should be no expectation that ODC operators will meet the IAU brightness recommendations.

Should ODCs be pursued, one could ask what is the least problematic design? 
There is no simple answer to this. 
On one hand, the use of multiple rings with multiple nodes increases satellite sky coverage.
This could suggest that a single ring is better than an X-ring. 
On the other hand, single rings could also make for denser sky structures.
Moreover, operators with different designs and nodal alignments, even if using single rings, could lead to high sky coverage.
The only clear way to reduce the on-sky impacts of ODCs is to limit their use.

It should be clear that there are numerous other impacts not discussed here. 
This paper has focused specifically on sky impacts of ODCs, and doing so is not meant to ignore those other concerns. 
Indeed, collision risks will be a major safety concern, and should be expected. 
Consequential meteoroid impacts will occur and affect operations, and other accidents, such as internal explosions and other failures will occur.
While there will always be risk with space operations, the current proposals have, in our view, substantial potential for setting the conditions for which near perfect operations will be needed, but unachievable. 

A reminder of this is the CRASH Clock, which is currently below three days \citep{Thiele2026}\footnote{See \url{https://outerspaceinstitute.ca/crashclock/} for the most up-to-date CRASH Clock value}.
The CRASH Clock is an environmental metric that asks what is the timescale for conjunctions to occur that could lead to a collision if all forms of satellite management were to stop. 
There is already little time to react to a widespread failure of a large system, and ODCs of tens of thousands to one million satellites could give essentially no possibility of having a usable reaction time.

ODCs, as proposed, could easily require in excess of 100-1000x the mass to LEO as exists now, and with this, a comparable increase in rocket exhaust. 
This will in turn lead to atmospheric pollution and chemical changes \citep[e.g.,][]{Shutler2022}.  
Moreover, a megaconstellation of one million satellites with 5 year operating lifetimes would require one disposal ever 3 minutes, resulting in metal deposition in the stratosphere orders-of-magnitude beyond anything yet modelled \citep[e.g.,][]{Maloney2025}.
And any imperfect, uncontrolled atmospheric disposal would result in serious ground casualty risks \citep[e.g.,][]{Wright2025}.

Earth is now a ringed planet with a debris system, with these features arising from human activity. 
The GEO satellite ring, its debris ring, other debris streams, and debris shells about Earth are tenuous and can still be managed. 
However, the continued rapid development and deployment of satellites is intensifying the anthropogenic debris system.
Earth's satellite and debris system is collisional \citep{Kessler1978}, a point that should not be taken lightly.

%\begin{acknowledgments}

This research has been supported in part by NSERC Discovery Grants RGPIN-2020-04111 (SML), RGPIN-2026-04644 (ACB) and RGPIN-2020-04513 (HR).
%\end{acknowledgments}

\software{This research was made possible by the open-source projects \texttt{REBOUND} \citep{Rein2012}, \texttt{WHFast} \citep{Rein2015},
\texttt{Jupyter} \citep{jupyter}, \texttt{iPython} \citep{ipython}, and \texttt{matplotlib} \citep{matplotlib, matplotlib2}.}

\bibliography{Bibliography}{}
\bibliographystyle{aasjournal}

\end{document}